\documentclass[preprint2]{aastex63}

\usepackage{natbib}

\usepackage{amsmath}
\usepackage{hyperref}
\usepackage{listings}
\usepackage{color}
\usepackage{float}
\usepackage{xspace} 
\usepackage{longtable}

\newcommand{\angstrom}{\mbox{\normalfont\AA}\xspace}
\newcommand{\kepler}{\textsl{Kepler}\xspace}
\newcommand{\logrprime}{\ensuremath{\log R^\prime_\mathrm{HK}}\xspace}
\newcommand{\moog}{\texttt{MOOG}\xspace}
\newcommand{\teff}{\ensuremath{T_{\mbox{\scriptsize eff}}}}
\newcommand{\logg}{\ensuremath{\log g}}
\newcommand{\vsini}{\ensuremath{v \sin i}}
\newcommand{\kms}{\ensuremath{\mbox{km s}^{-1}}}
\newcommand{\gaia}{\textit{Gaia}\xspace}

\begin{document}

\title{Stellar Properties of Active G and K Stars: Exploring the Connection Between Starspots and Chromospheric Activity}

\author[0000-0003-2528-3409]{Brett M. Morris}
\affiliation{Astronomy Department, University of Washington, Seattle, WA 98195, USA}

\author[0000-0002-2792-134X]{Jason L. Curtis}
\altaffiliation{NSF Astronomy and Astrophysics Postdoctoral Fellow}
\affiliation{Department of Astronomy, Columbia University, 550 West 120th Street, New York, NY 10027, USA}

\author[0000-0002-5095-4000]{Charli Sakari}

\author[0000-0002-6629-4182]{Suzanne L. Hawley}
\affiliation{Astronomy Department, University of Washington, Seattle, WA 98195, USA}

\author[0000-0002-0802-9145]{Eric Agol}
\altaffiliation{Guggenheim Fellow}
\affiliation{Astronomy Department and Virtual Planetary Laboratory, University of Washington, Seattle, WA 98195, USA}

\email{morrisbrettm@gmail.com}

\begin{abstract}
We gathered high resolution spectra for an ensemble of 55 bright active and inactive stars using the ARC 3.5 m Telescope Echelle Spectrograph at Apache Point Observatory ($R\approx$31,500). We measured spectroscopic effective temperatures, surface gravities and metallicities for most stars in the sample with SME and MOOG. Our stellar property results are consistent with the photometric effective temperatures from the \textit{Gaia} DR2 pipeline. We also measured their chromospheric $S$ and \logrprime indices to classify the stars as active or inactive and study the connection between chromospheric activity and starspots.  We then attempted to infer the starspot covering fractions on the active stars by modeling their spectra as a linear combination of hot and cool inactive stellar spectral templates. We find that it is critical to use precise colors of the stars to place stringent priors on the plausible spot covering fractions. The inferred spot covering fractions generally increase with the chromospheric activity indicator \logrprime, though we are largely insensitive to spot coverages smaller than $f_S \lesssim 20$\%. We find a dearth of stars with small \logrprime and significant spot coverages.  
\end{abstract}

\section{Introduction}

One of the many remarkable data products produced by the ESA \gaia mission is a catalog of stellar effective temperatures \citep{GaiaDR2}. The \gaia Apsis pipeline uses two-color photometry and parallaxes to measure effective temperatures for a broad range of stars and other objects, totalling more than 10$^8$ sources \citep{Bailer-Jones2013, DR2prop}. This indispensable resource will be a touchstone catalog for years to come, and as such, it is important to validate its results on bright, nearby stars for which the effective temperatures may be unambiguously determined using other means. In this work, we use high resolution spectroscopy analyzed with two independent methods to measure effective temperatures, and compare results to the ESA \gaia effective temperatures.

We also used our spectroscopy to investigate stellar activity. In a high resolution spectrum, starspots change a star's flux as a function of \textit{wavelength}, as well as time. Magnetic starspots are cooler than the surrounding photosphere, and to a first approximation, they emit the spectrum of a cooler star. These starspots are the observable, outer manifestations of the stellar dynamos operating near and below the surface \citep{Berdyugina2005}. 

Magnetically active regions on stellar surfaces play an increasingly important role in understanding exoplanets, both with radial-velocity measurements and transit-transmission spectroscopy \citep[][and references therein]{Cameron2017, Zellem2017}.  Uncorrected variations in the stellar flux in time and wavelength can cause degeneracies when inferring the atmospheric properties of a transiting exoplanet, which can lead to, for example, false positive detections of Rayleigh scattering (see \citealt{Sing2011, McCullough2014, Oshagh2014}), or  transparent atmospheres in blue spectral regions \citep{Rackham2018, Morris2018c}.
Direct measurement of starspot covering fractions are thus important inputs for models of photometry and spectra of active stars, which will be used to measure the transmission spectra of exoplanet atmospheres \citep{Wakeford2019}, exoplanet transit timing variations, and host star radial velocities.  

The full covering fraction of starspots can be difficult to infer from their rotational variability. Only the time-varying, low-order spherical harmonics contribute to the total flux variability; stationary spots or odd harmonics yield zero net variability, while time-variable high-order even harmonics are strongly suppressed.  Thus, longitudinally- or symmetrically distributed spots or slowly evolving spots on stars with long rotational periods can yield an underestimate of the total spot covering fraction. Hence, spectroscopy may be a preferable approach for measuring the total covering fraction of starspots.

One effort towards direct measurements of spot temperatures and covering fractions outside of the Solar System was led by \citet{Neff1995} \citep[see also][]{oneal1996, oneal1998, ONeal2001, ONeal2004}. The authors modeled echelle spectra of active stars with a linear combination of model stellar atmospheres from a G dwarf and an M giant. In particular, they found that titanium oxide (TiO) absorption at 7054 and 8859 \angstrom due to starspots yielded constraints on the spot temperatures and spot covering fractions. The TiO molecule is a useful tracer for starspots as it forms below temperatures of $4000$ K---cool starspots near or below this temperature will produce some TiO absorption, while the rest of the photosphere of a G or K star ($\teff > 4000$ K) will not.

We present a library of stellar template spectra of more than 50 stars taken with the Astrophysical Research Consortium Echelle Spectrograph (ARCES) on the ARC 3.5 m Telescope at Apache Point Observatory in Section~\ref{sec:obs}. For each star we compute effective temperatures with a variety of techniques, and measure the activity indices $S$ and $\log R^\prime_\mathrm{HK}$. In Section~\ref{sec:tio} we model the spectra of a subset of the active stars with linear combinations of the inactive stellar template spectra to infer the spot area coverage on the active stars. We discuss the results and conclusions in Sections~\ref{sec:discuss} and \ref{sec:conclusion}.

\section{Stellar Parameters} \label{sec:obs}

\subsection{Sample Selection}

We selected bright, northern targets observable from Apache Point Observatory for inclusion in this work. We chose only stars with previously measured $S$-indices, a measurement of chromospheric emission indicative of magnetic activity  \citep{Wilson1968}, then divided the sample into two classes: stars that are chromospherically active or inactive. 

The stellar sample of 55 stars is comprised of 29 active stars and 26 inactive stars, for a total of one F, 18 G, 34 K, and two M stars. The stellar sample is listed in Table~\ref{tab:megatable}, and detailed descriptions of each target when available are given in Appendix~\ref{app:stellarprops}. 

We selected a set of stars with low-value $S$-indices to use as templates assuming stars are essentially spotless. We also selected a set of active stars spanning the full dynamic range of the $S$-index. In the next section, we will fit the spectra of the active stars as a linear combinations of spectra of inactive stars (with small $S$-indices).

\subsection{Raw spectrum reduction} 

\begin{figure}
    \centering
    \includegraphics[scale=0.85]{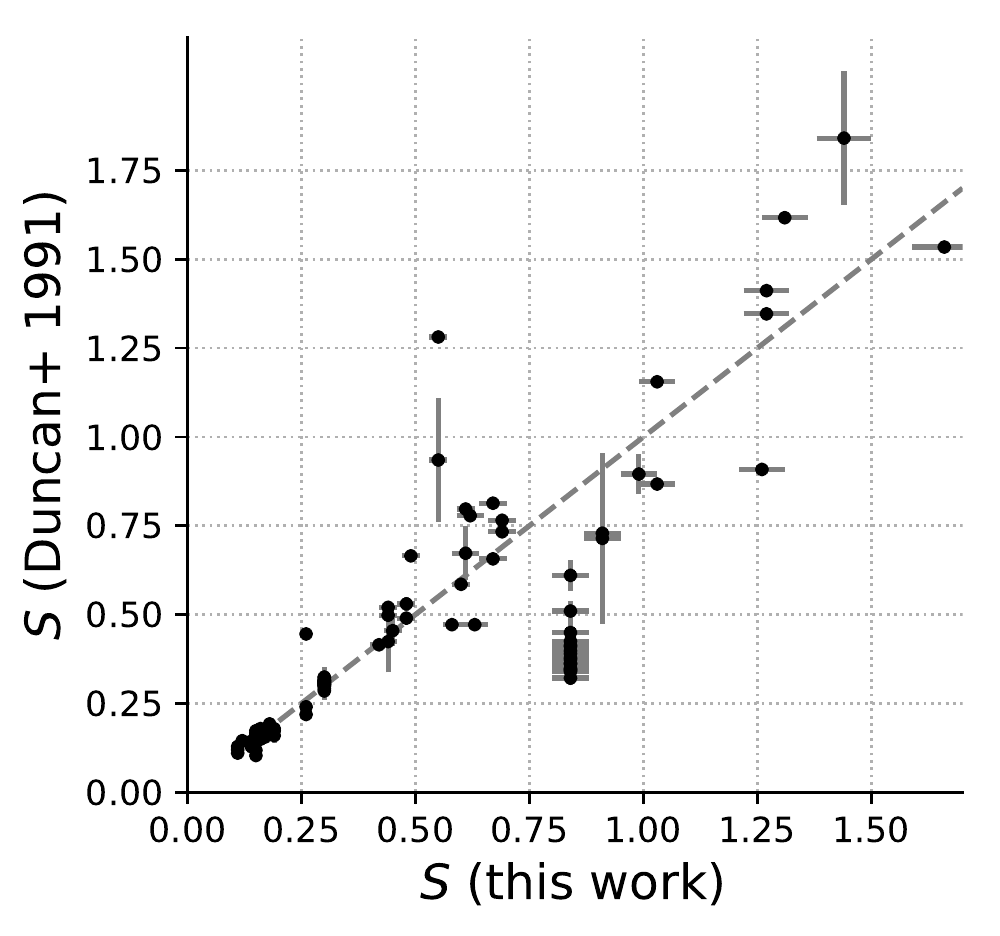}
    \caption{$S$-index measured in this work, compared with previous Mount Wilson Observatory (MWO) observations by \citet{Duncan1991}.}
    \label{fig:sind_correlation}
\end{figure}

\begin{figure}
\centering
\includegraphics[scale=0.8]{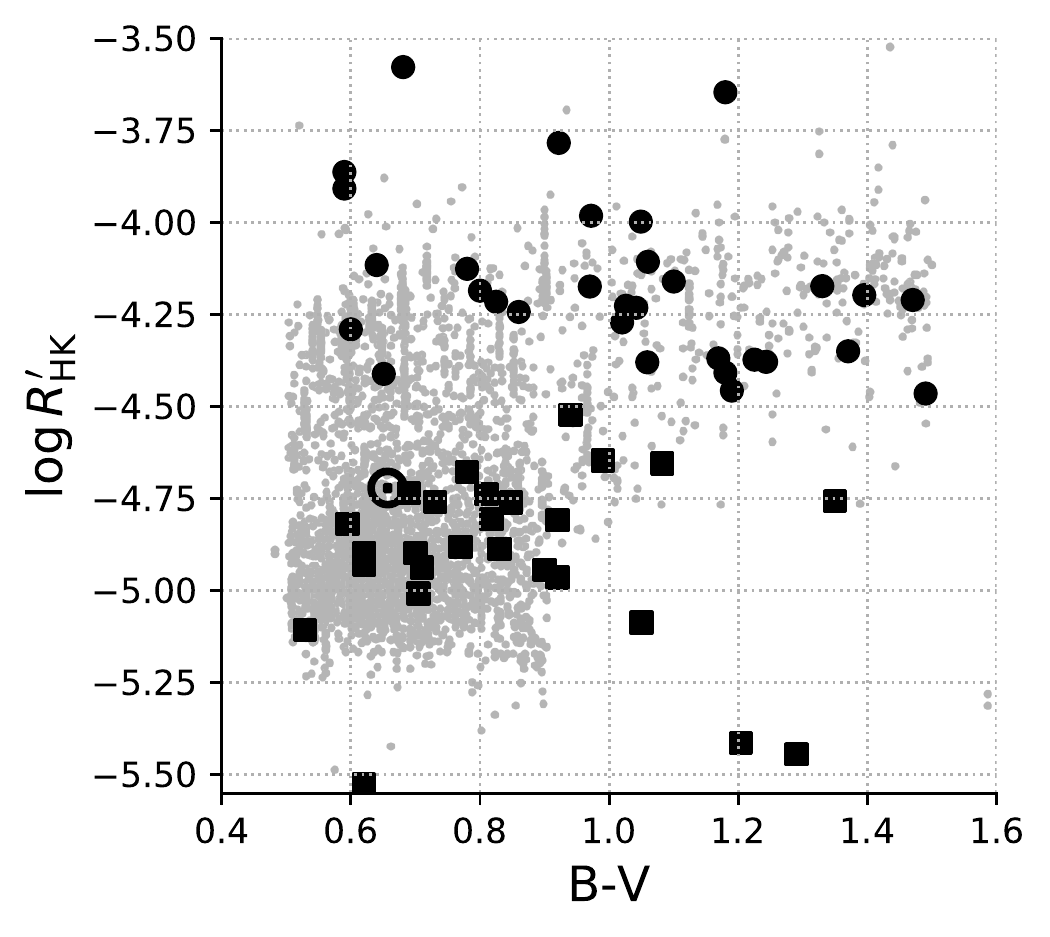}
\caption{Chromospheric activity \logrprime index as a function of $B-V$ color for our template stars (black), compared with the large field samples of \citet{Duncan1991}, \citet{Wright2004} and \citet{Isaacson2010}, (gray) computed with the procedure outlined in \citet{Mittag2013}. Stars classified as ``inactive'' are marked with black squares, stars classified as ``active'' are marked with black circles. The outliers at low activity are giants, including G giant HD 5857, and the K giants Gl 705, HD 6497 and HD 145742 (see Table~\ref{tab:megatable} for stellar properties).}
\label{fig:rhk}
\end{figure}

\begin{figure}
    \centering
    \includegraphics[scale=0.6]{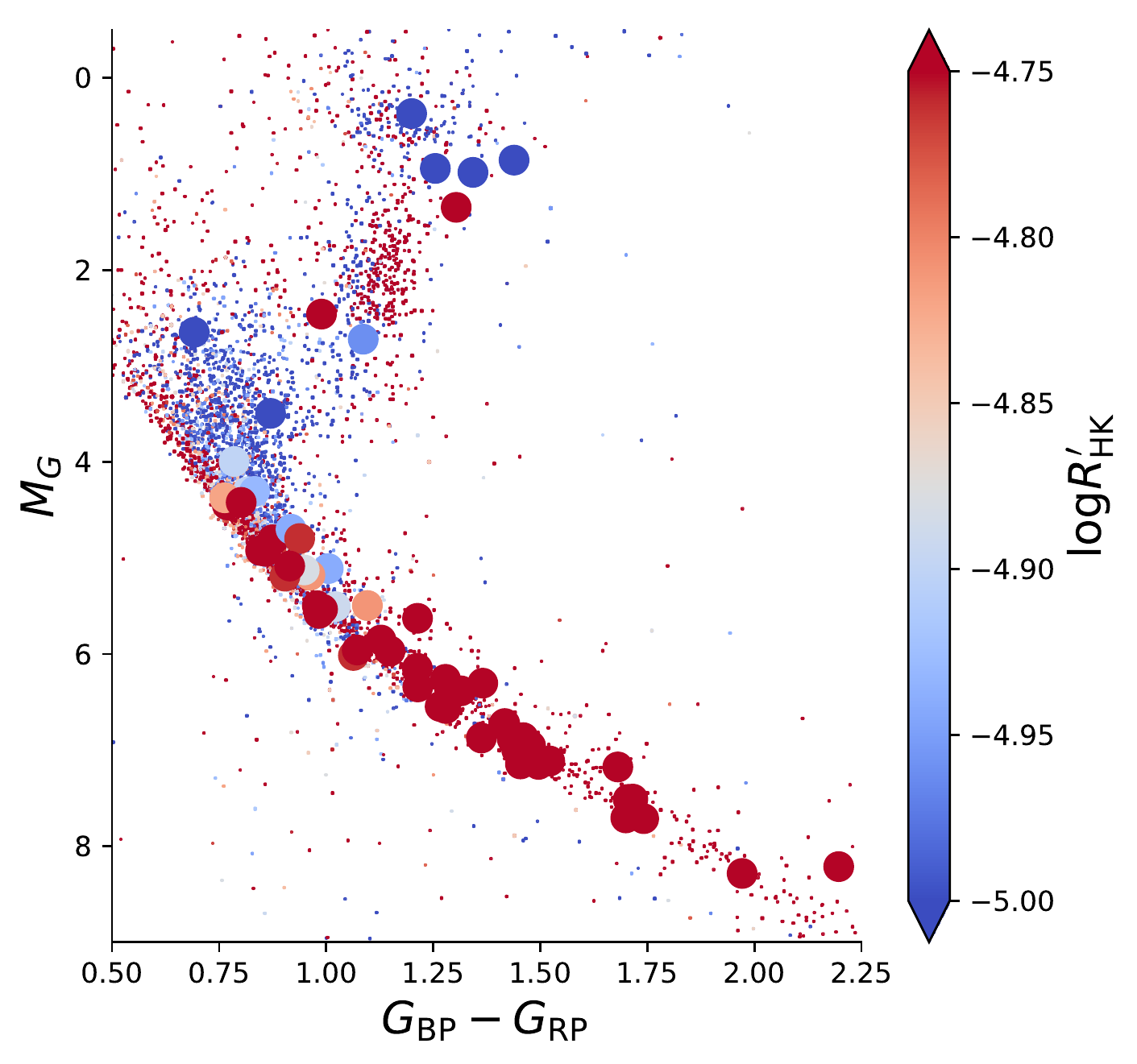}
    \caption{Color--magnitude diagram from \citet{GaiaDR2}, colored by \logrprime, for stars in \citet{Duncan1991}, \citet{Wright2004} and \citet{Isaacson2010} (small points). The larger circles are the stars observed in this work.}
    \label{fig:cmd}
\end{figure}

We reduced the raw ARCES spectra with \texttt{IRAF} methods to subtract biases, removed cosmic rays, normalized by the flat field, and carried out the wavelength calibration with exposures of a thorium-argon lamp.\footnote{An ARCES data reduction manual by J. Thorburn is available at \url{http://astronomy.nmsu.edu:8000/apo-wiki/attachment/wiki/ARCES/Thorburn_ARCES_manual.pdf}}
We fit each order of the spectrum of an early-type star with a high-order polynomial to measure the blaze function, and divided the spectra of the target stars by that polynomial to normalize each spectral order.

Next, the normalized spectra must be shifted in wavelength into the rest-frame by removing their radial velocities. We removed the radial velocity by maximizing the cross-correlation of the ARCES spectra with PHOENIX model atmosphere spectra \citep{Husser2013}. We fit a second-order polynomial with robust least-squares (with a Cauchy loss function) to the continuum in each order, before stitching together neighboring orders, taking the mean flux in regions of overlapping wavelength coverage. The source code for the pipeline used for wavelength corrections and continuum normalization is available online \citep{aesop}.

\subsection{Activity classification}

We calculated the chromospheric \ion{Ca}{2} H \& K activity $S$ and \logrprime indices by following the algorithm outlined in \citet{Morris2017b}. In summary, we computed the flux within triangular-weighted windows centered on the \ion{Ca}{2} H \& K features, normalized by pseudo-continuum to the red and blue of the H and K features. We then translated the instrument-native $S_{APO}$ index into the $S_{MWO}$ index for comparison with data collected from other instruments. Figure~\ref{fig:sind_correlation} compares our $S$-indices with measurements for a subset of stars from the Mount Wilson Observatory HK Project \citep{Duncan1991}, and we find good agreement. With these $S$ indices, we follow the procedure outlined in \citet{Mittag2013} for computing \logrprime. The results are shown in Figure~\ref{fig:rhk} as a function of stellar $(B - V)$ color, 
and in Figure~\ref{fig:cmd} on the \gaia DR2 color--magnitude diagram (CMD), 
and are listed in Table~\ref{tab:megatable}. Figure~\ref{fig:rhk} illustrates where we have drawn our cutoff between active and inactive stars, and Figure~\ref{fig:cmd} demonstrates that it is possible to distinguish the evolutionary states of these bright stars from Gaia photometry and parallaxes alone, even before we apply high resolution spectroscopy.

\begin{figure}
    \centering
    \includegraphics[scale=1]{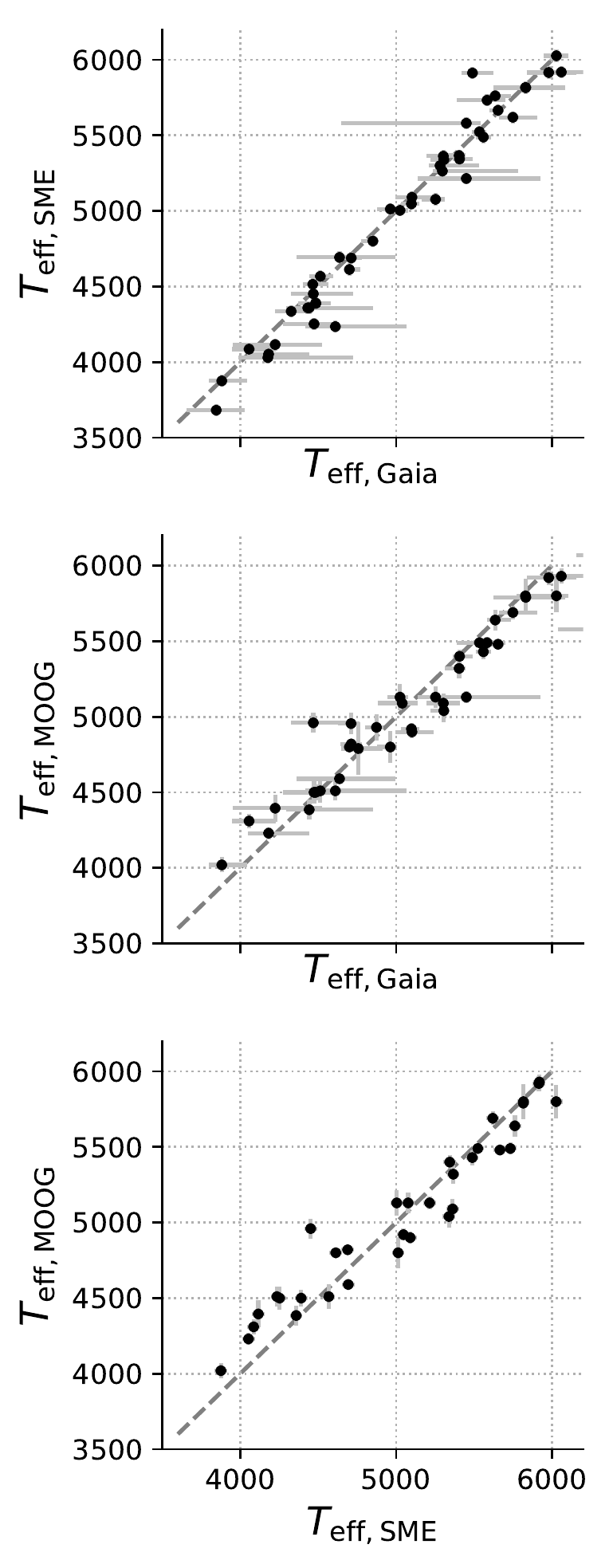}
    \caption{Comparisons between effective temperatures [K] computed with SME, \moog and from the {\it Gaia} DR2 catalog.}
    \label{fig:temps}
\end{figure}

\subsection{Stellar parameters}

In this section we measure the effective temperatures of the stars with three techniques: using the ARCES high resolution spectroscopy independently with SME and \moog, and using \gaia DR2 broadband photometry \citep{DR2prop}. We estimated the effective temperatures with three techniques so that we could determine if activity was affecting the \teff\ estimates. In the following sections we will outline the effective temperature measurement with each technique, and compare the results to one another in Section~\ref{sec:teffcomparison}.

\subsubsection{Stellar parameters with SME} \label{sec:sme}

We used Spectroscopy Made Easy \citep[SME;][]{sme, Valenti2005, Piskunov2017} version 522 to compute the effective temperatures (\teff), surface gravities (\logg),  metallicities ([Fe/H]), and projected rotational velocities (\vsini) of all stars in our sample. 
The SME algorithm fits selected spectral regions with synthetic spectra 
generated with a particular atmospheric model (ATLAS12; \citealt{atlas12}) and line list. 
We followed the basic procedure and adopted the wavelength range and line list from \citet{Valenti2005}, 
and analyzed the 5164-5190 \AA\ interval covering the pressure-sensitive \ion{Mg}{1} $b$ triplet, 
and the 6000-6070 and 6100-6180 \AA\ intervals split into 7 segments.

We normalized the \ion{Mg}{1} $b$ segment by comparing a solar spectrum (observed with ARCES, reflected off the Moon) 
to the \citet{Wallace2011} solar atlas, where we matched the velocity offset and instrumental broadening 
between spectra, then fit a polynomial to the ratio of the two spectra to derive a blaze model for this order.
For the other spectral segments, this procedure also works, but the absorption lines in the regions 
are sparse enough that polynomials fit directly to the spectra are also effective. 
All spectra were cross-correlated against the solar atlas during normalization. 
SME also fits for small radial velocity offsets and residual linear trends across spectral segments. 

We initialized SME with properties taken from the literature; when spectroscopic results were not 
available, we used \gaia DR2 \teff\ values; 
when other parameters were unavailable, we adopted 
\logg\ = 4.6~dex, [Fe/H] = 0.0 dex, and \vsini\ = 2 \kms.
Following \citet{Valenti2005}, 
microturbulence was fixed to 0.85~\kms\ and we
adopted their macroturbulence relation. 
We did not solve for any additional abundances besides the metallicity parameter for this work,
which therefore assumes a scaled-solar composition.
After running the initial fit, 
we perturbed the resulting \teff\ by $\pm200$ K and refit the spectra. 
Fits typically converged to within 10 K of the first solution for 
stars warmer than $\teff > 4600$ K.

The spectra of cooler stars are more complicated, particularly in 
the \ion{Mg}{1} $b$ order. 
For stars cooler than $\teff < 4600$ K,
we removed this order from our analysis, 
then fixed \logg\ to a value drawn from a 3 Gyr (i.e., middle-aged) Dartmouth isochrone with [Fe/H] = +0.10 dex 
(this choice was arbitrary and made purely for our convenience) 
at the initial \teff , 
and iterated the fit once.
We verified convergence by  
varying the initial \teff\ by $\pm$500 K, 
and resetting values each time to [Fe/H] = 0.0 dex and \logg\ = 4.6 dex.

\citet{Ryabchikova2016} assessed the stellar parameter accuracies attainable with SME 
for FGK dwarfs with high-quality spectra (e.g., $S/N = 200$, $R = $40,000-110,000), 
and found uncertainties on
\teff\ of 50-70 K, 
on \logg\ of 0.1 dex,
and 
on metallicity of $0.05-0.06$ dex \citep[see also][]{Piskunov2017, Brewer2016}.
While their spectroscopy methodology (including wavelength regions analyzed) 
differs from ours, 
these values provide error floors that are consistent with our results.
Results for stars with multiple observations and results from varying the initial conditions 
typically yielded dispersions much lower than this error floor.
For now, we will assume that our uncertainties for 
\teff\ are 50 K, and will later compare our values to 
the photometric \textit{Gaia} DR2 and literature values to 
assign final uncertainties. The results are shown in Figure~\ref{fig:temps}.

\subsubsection{Stellar parameters with \moog} \label{sec:moog}
 
For most of the stars, \teff, \logg, microturbulent velocity ($\xi$, in km s$^{-1}$), and [Fe/H] were also determined with a standard model atmospheres approach.  Equivalent width (EWs) of \ion{Fe}{1} and \ion{Fe}{2} lines were first measured in the normalized spectra using the Gaussian profile fitting code {\tt DAOSPEC} \citep{DAOSPECref}.  {\tt DAOSPEC} has difficulty measuring lines with $\rm{EW} > 100$ m\AA; as a result, those lines were remeasured by hand.  The lines from \citet{Fulbright2006} and \citet{McWilliam2013} were utilized.  Lines blueward of 5000 \AA $;$ were not included to avoid issues with continuum normalization. Strong lines (with reduced EWs\footnote{$\rm{REW} = \log(\rm{EW}/\lambda)$, where $\lambda$ is the wavelength of the transition} $REW>-4.8$) were not included, to minimize uncertainties due to, e.g., damping constants \citep{McWilliam1995}.

The EWs were then input into the 2017 version of the Local Thermodynamic Equilibrium (LTE) line analysis code {\tt MOOG} \citep{Sneden}.  ATLAS model atmospheres were adopted \citep{KuruczModelAtmRef}, using an interpolation scheme for temperatures and surface gravities that lie between the grid points. For each Fe line, [Fe/H] was calculated relative to the solar abundance for that line (using the solar EWs from \citealt{Fulbright2006} and \citealt{McWilliam2013}).  The appropriate \teff\ and  $\xi$ were determined by minimizing trends in [\ion{Fe}{1}/H] with excitation potential (EP, in eV) and REW, respectively.  The appropriate \logg was determined with the BaSTI isochrones \citep{BaSTIref}, utilizing the spectroscopic \teff.  This means that the surface gravity could be mildly dependent on the adopted age and metallicity; however, the effect on main sequence stars is small.  The [\ion{Fe}{1}/H] and [\ion{Fe}{2}/H] ratios are averages from the individual lines.  The surface gravities were {\it not} determined by requiring ionization equilibrium (i.e., by requiring that the \ion{Fe}{1} and \ion{Fe}{2} abundances be equal), both due to larger uncertainties in the \ion{Fe}{2} abundances and possible non-LTE effects \citep{KraftIvans2003}.  While there are $\sim 10-120$ acceptable \ion{Fe}{1} lines per star (depending on metallicity, \teff, and S/N), there are many fewer \ion{Fe}{2} lines; as a result, the random errors in [\ion{Fe}{2}/H] are much larger than those in [\ion{Fe}{1}/H].

Note that none of these parameters are independent; the adopted temperatures are therefore dependent upon the other parameters, particularly the microturbulent velocity.  The errors in \teff were determined by assuming errors in $\xi$ of $\pm 0.3$ km s$^{-1}$ (errors in \logg and [Fe/H] have a negligible effect on the temperatures).

The \moog effective temperatures are listed alongside the others in Table~\ref{tab:megatable}; the rest of the \moog results ($\log g$, [\ion{Fe}{1}/H], [\ion{Fe}{2}/H], and microturbulence) are enumerated in Appendix~\ref{sec:moogresults}.  The temperatures generally agree with the {\it Gaia} values, though they may be slightly higher in the coolest stars; see Figure~\ref{fig:temps}.  The cooler stars have fewer suitable \ion{Fe}{1} lines for an EW analysis; as a result, the temperatures are likely to be more sensitive to line-to-line variations as a result of, e.g., continuum placement or atomic data.

\subsubsection{Stellar temperatures with ESA/\gaia DR2} 
\label{sec:dr2}
The European Space Agency's \gaia mission recently delivered 
astrometric, photometric, and radial velocity data for 
a large number of stars \citep{GaiaDR2}. 
The second data release (DR2) included an analysis of the three-band photometry ($G_{\rm BP}, G_{\rm RP}, G$), 
with the Apsis pipeline \citep{apsis2013}, 
and derived effective temperatures for 161 million sources brighter than $G<17$ 
with $3000 < \teff < 10000$ K \citep{DR2prop}. 
The Apsis pipeline consists of a machine learning algorithm trained on 
large \teff\ data sets from the literature (e.g., LAMOST, RAVE, the \textit{Kepler} Input Catalog), 
and was validated using 
data from GALAH, nearby solar twin stars, and other sources.

We performed additional tests to assess the accuracy and 
precision of the DR2 catalog effective temperatures by comparing results 
for well-characterized nearby stars, 
including the FGK stars observed by CPS with Keck/HIRES and characterized with SME by \citet{Brewer2016}, 
the K and M dwarfs characterized with interferometry and bolometric fluxes by 
\citet{Boyajian2012}, 
and the M dwarfs characterized with optical and NIR spectroscopy by \citet{Mann2015}.
The \textit{Gaia} DR2 \teff\ values are consistent with this joint benchmark sample
for $\teff > 4100$ K, but diverge at cooler temperatures.
The warmer stars show remarkable agreement, having an rms of only 70 K, which is 
much lower than the typical \teff\ accuracy of 324 K quoted by \citet{DR2prop}.
This is due to the fact that these stars are all nearby and likely appear unreddened.
We fit a polynomial to the DR2 color $(G_{\rm BP} - G_{\rm RP})$ versus 
the benchmark sample \teff\ values to re-calibrate the DR2 photometric temperatures 
and place the cool stars on the benchmark sample's temperature scale.
We will use this re-calibration later in the next subsection.
Table \ref{tab:megatable} reports the DR2/Apsis values \citep{DR2prop}.

\subsection{Comparison of stellar temperatures} \label{sec:teffcomparison}

Figure~\ref{fig:temps} compares the effective temperatures we derived with SME and \moog with the \gaia DR2 effective temperatures. 
The SME and \moog temperatures have significantly better precision than the \gaia temperatures; 
however, in many cases, it is possible that \citet{DR2prop} overestimated the uncertainties---we reiterate that 
remarkable agreement is found when comparing \teff\ values for benchmark FGKM dwarfs from 
\citet{Brewer2016}, \citet{Boyajian2012}, and \citet{Mann2015} with DR2,  with an rms of only 70 K.
In fact, many of the stars in our sample are drawn from those three catalogs.

Of the 55 stars in our sample, 19 are also in \citet{Brewer2016}, five are in SPOCS \citep{Valenti2005}, one is in \citet{Boyajian2012}, and two are in \citet{Mann2015}. 
The SB2 spectrum of HD 45088 was characterized by \citet{Glazunova2014}, and 
61 Cyg A and B are \textit{Gaia} FGK benchmark stars \citet{Heiter2015}. 
This means that 30 of 55 stars, over half of our sample, have been precisely characterized by these other studies.
The median and standard deviation of the difference between our SME values and the benchmark values is 6 $\pm$ 74 K, where our stars are negligibly cooler, systematically.
This rms drops to 43 K when considering the 22 stars with benchmark $\teff > 4800$ K, 
and increases to 85 K when considering only the 7 cooler stars (the SB2 is not included in this analysis).
Based on this consistency, we 
set the SME uncertainties to 40 K for stars warmer than $\teff = 4800$ K, and 100 K for stars cooler than $\teff = 4800$ K.
Unless otherwise noted below, we choose to use the SME effective temperatures because: (1) they show better agreement with the \gaia effective temperatures, and (2) the SME analysis converged on effective temperatures for the largest number of the stars in the sample, compared to our MOOG analysis.

\begin{longrotatetable}
\startlongtable
\begin{deluxetable*}{l|c|c|ccc|ccc|cc}
\tablecaption{Spectral types, Stellar colors, effective temperatures, SME properties ($\log g$, [Fe/H], $v \sin i$), $S$-indices and \logrprime indices.Stars with names marked with $^*$ are listed as BY Dra variables; $^\dagger$ denotes long-period binary stars; $^\ddagger$ denotes RS CVn binaries. \label{tab:megatable}}
\tablehead{\colhead{Target} & \colhead{Sp.T.} & \colhead{$G_\mathrm{BP} - G_\mathrm{RP}$} & \colhead{$T_\mathrm{eff,Gaia}$} & \colhead{$T_\mathrm{eff, SME}$} & \colhead{$T_\mathrm{eff, MOOG}$} & \colhead{$\log g$} & \colhead{[Fe/H]} & \colhead{$v \sin i$} & \colhead{$S$} & \colhead{$\log R^\prime_\mathrm{HK}$} }
\startdata
\sidehead{\textit{Active}} 
HD 151288 & K7.5Ve & 1.7053 & $4224^{+300}_{-270}$ & $4337 \pm 100$ & $4395 \pm 90$ & 4.68 & -0.117 & 3.64 & $1.27 \pm 0.05$ & $-3.58 \pm 0.02$ \\
Gl 702 B & K4V & 1.4774 & $4442^{+408}_{-150}$ & $4359 \pm 100$ & $4385 \pm 65$ & 4.65 & -0.144 & 2.06 & $0.84 \pm 0.04$ & $-3.61 \pm 0.02$ \\
EQ Vir$^*$ & K5Ve & 1.496 & $4433^{+82}_{-106}$ & $4359 \pm 100$ &  & 4.65 & -0.144 & 2.06 & $3.00 \pm 0.12$ & $-3.65 \pm 0.02$ \\
HD 175742$^\ddagger$ & K0V & 1.213 & $4874^{+43}_{-74}$ & $4980 \pm 40$ & $4930 \pm 85$ & 4.4 & -0.09 & 14.7 & $1.26 \pm 0.05$ & $-3.78 \pm 0.02$ \\
EK Dra & G1.5V & 0.875 & $5584^{+115}_{-193}$ &  &  &  &  &  & $0.58 \pm 0.02$ & $-3.91 \pm 0.02$  \\
Kepler-63 & G2V & 0.9149 & $5450^{+92}_{-805}$ & $5581 \pm 40$ &  & 4.69 & 0.129 & 5.43 & $0.45 \pm 0.02$ & $-3.95 \pm 0.03$ \\
HD 45088$^*$ & K2Ve & 1.2134 & $4786^{+39}_{-36}$ & $5025 \pm 40$ &  &  &  &  & $0.91 \pm 0.04$ & $-3.98 \pm 0.02$ \\
HD 113827 & K4V & 1.4537 & $4610^{+459}_{-191}$ & $4235 \pm 100$ & $4510 \pm 65$ & 4.66 & -0.381 & 2.59 & $1.03 \pm 0.04$ & $-4.00 \pm 0.02$ \\
HD 134319$^*$ & G5V & 0.8469 & $5636^{+100}_{-54}$ & $5762 \pm 40$ & $5640 \pm 70$ & 4.59 & -0.01 & 10.9 & $0.42 \pm 0.02$ & $-4.12 \pm 0.02$ \\
HD 127506 & K3.5V & 1.2794 & $4711^{+55}_{-83}$ & $4651 \pm 100$ & $4955 \pm 70$ & 4.78 & -0.265 & 3.32 & $0.49 \pm 0.02$ & $-4.13 \pm 0.02$ \\
HD 200560 & K2.5V & 1.1492 & $4963^{+42}_{-85}$ & $5013 \pm 40$ & $4800 \pm 105$ & 4.66 & 0.09 & 3.6 & $0.61 \pm 0.03$ & $-4.17 \pm 0.02$ \\
HD 88230 & K6VeFe-1 & 1.7162 & $4057^{+174}_{-107}$ & $4086 \pm 100$ & $4310 \pm 50$ & 4.68 & -0.167 & 0.1 & $1.31 \pm 0.05$ & $-4.17 \pm 0.02$ \\
HD 220182$^*$ & G9V & 0.9794 & $5303^{+106}_{-106}$ & $5363 \pm 40$ & $5090 \pm 65$ & 4.66 & 0.0 & 4.6 & $0.45 \pm 0.02$ & $-4.19 \pm 0.02$ \\
HD 266611 & K5V & 1.7415 & $4182^{+260}_{-132}$ & $4053 \pm 100$ & $4230 \pm 40$ & 4.69 & -0.296 & 0.1 & $1.44 \pm 0.06$ & $-4.20 \pm 0.02$ \\
HD 209290 & M0.5V & 1.9714 & $3882^{+163}_{-82}$ & $3876 \pm 100$ & $4020 \pm 50$ & 4.72 & -0.321 & 2.68 & $1.66 \pm 0.07$ & $-4.21 \pm 0.02$ \\
HD 41593$^*$ & K0V & 0.9916 & $5306^{+28}_{-83}$ & $5339 \pm 40$ & $5040 \pm 75$ & 4.64 & -0.01 & 2.4 & $0.44 \pm 0.02$ & $-4.22 \pm 0.02$ \\
HD 82106 & K3V & 1.214 & $4749^{+71}_{-40}$ & $4726 \pm 100$ &  & 4.45 & -0.06 & 0.4 & $0.61 \pm 0.02$ & $-4.23 \pm 0.02$ \\
HD 79555$^\dagger$ & K4V & 1.3148 & $4861^{+149}_{-251}$ & $4744 \pm 100$ &  & 4.74 & -0.19 & 0.1 & $0.62 \pm 0.03$ & $-4.23 \pm 0.02$ \\
HD 87884$^\dagger$ & K0Ve & 1.0725 & $5098^{+50}_{-69}$ & $5047 \pm 40$ & $4920 \pm 40$ & 4.63 & -0.25 & 2.5 & $0.44 \pm 0.02$ & $-4.24 \pm 0.02$ \\
HD 222107$^\ddagger$ & G8IVk & 1.3037 & $4637^{+360}_{-274}$ & $4693 \pm 100$ & $4590 \pm 30$ & 3.08 & -0.41 & 7.8 & $0.55 \pm 0.02$ & $-4.27 \pm 0.02$ \\
HD 39587$^\ddagger$ & G0V & 0.7691 & $6028^{+76}_{-81}$ & $6027 \pm 40$ & $5800 \pm 110$ & 4.58 & -0.01 & 8.8 & $0.30 \pm 0.01$ & $-4.29 \pm 0.03$ \\
61 Cyg B$^\dagger$ & K7V & 1.6997 & $4177^{+548}_{-189}$ & $4030 \pm 100$ &  & 4.69 & -0.413 & 0.1 & $0.97 \pm 0.04$ & $-4.35 \pm 0.02$ \\
HD 149957 & K5V & 1.4601 & $4485^{+37}_{-89}$ & $4389 \pm 100$ &  & 4.65 & -0.007 & 4.54 & $0.67 \pm 0.03$ & $-4.37 \pm 0.02$ \\
HD 122120 & K5V & 1.4146 & $4469^{+253}_{-144}$ & $4452 \pm 100$ & $4960 \pm 65$ & 4.64 & -0.132 & 1.19 & $0.60 \pm 0.02$ & $-4.37 \pm 0.02$  \\
HD 47752 & K3.5V & 1.2662 & $4700^{+69}_{-22}$ & $4613 \pm 100$ & $4800 \pm 30$ & 4.63 & -0.306 & 0.1 & $0.48 \pm 0.02$ & $-4.38 \pm 0.02$ \\
HD 148467 & K6V & 1.5221 & $4474^{+153}_{-201}$ & $4253 \pm 100$ & $4500 \pm 75$ & 4.66 & -0.321 & 0.1 & $0.69 \pm 0.03$ & $-4.38 \pm 0.02$ \\
61 Cyg A$^\dagger$ & K5V & 1.445 & $4327^{+101}_{-103}$ & $4336 \pm 100$ &  & 4.65 & -0.493 & 0.65 & $0.57 \pm 0.02$ & $-4.41 \pm 0.02$ \\
HAT-P-11 & K4V & 1.2787 & $4757^{+154}_{-111}$ & $4800 \pm 100$ & $4790 \pm 175$ &  &  &  & $0.53 \pm 0.02$ & $-4.46 \pm 0.02$ \\
GJ 9781 A$^\dagger$ & K7 & 1.4343 & $4484^{+95}_{-110}$ & $4391 \pm 100$ & $4500 \pm 55$ & 4.65 & -0.239 & 0.1 & $0.99 \pm 0.04$ & $-4.46 \pm 0.02$ \\
\sidehead{\textit{Inactive}} 
HD 110833$^\dagger$ & K3V & 1.128 & $5025^{+50}_{-82}$ & $5004 \pm 40$ & $5130 \pm 85$ & 4.64 & 0.06 & 0.5 & $0.30 \pm 0.01$ & $-4.52 \pm 0.02$ \\
HR 8832 & K3V & 1.2078 & $4787^{+92}_{-73}$ & $4828 \pm 40$ &  & 4.63 & 0.03 & 0.2 & $0.26 \pm 0.01$ & $-4.65 \pm 0.03$ \\
GJ 4099 & M2.0Ve & 2.1969 & $3846^{+181}_{-190}$ & $3682 \pm 100$ &  & 4.77 & -0.494 & 5.68 & $0.87 \pm 0.05$ & $-4.67 \pm 0.02$  \\
$\sigma$ Draconis & K0V & 0.983 & $5450^{+474}_{-314}$ & $5215 \pm 40$ & $5130 \pm 40$ & 4.58 & -0.28 & 0.1 & $0.20 \pm 0.01$ & $-4.68 \pm 0.03$ \\
HD 68017$^\dagger$ & G3V & 0.8587 & $5749^{+155}_{-90}$ & $5620 \pm 40$ & $5690 \pm 45$ & 4.76 & -0.37 & 0.1 & $0.18 \pm 0.01$ & $-4.73 \pm 0.04$ \\
HD 67767 & G7V & 0.9891 & $5297^{+486}_{-61}$ & $5264 \pm 40$ &  & 3.81 & -0.1 & 0.1 & $0.19 \pm 0.01$ & $-4.74 \pm 0.03$ \\
HD 73667 & K2V & 1.0636 & $5039^{+98}_{-159}$ & $4800 \pm 100$ & $5090 \pm 30$ & 4.38 & -0.68 & 0.1 & $0.19 \pm 0.01$ & $-4.76 \pm 0.03$ \\
Kepler-17 & G2V & 0.9381 & $5490^{+132}_{-72}$ & $5913 \pm 40$ &  & 4.72 & 0.381 & 5.05 & $0.40 \pm 0.02$ & $-4.76 \pm 0.03$  \\
HD 62613 & G8V & 0.9037 & $5560^{+47}_{-49}$ & $5489 \pm 40$ & $5430 \pm 50$ & 4.53 & -0.12 & 0.1 & $0.18 \pm 0.01$ & $-4.76 \pm 0.04$ \\
HD 14039 & K1V & 1.0963 & $5072^{+98}_{-73}$ &  &  &  &  &  & $0.19 \pm 0.01$ & $-4.81 \pm 0.03$   \\
HD 182488 & K0V & 0.9621 & $5404^{+41}_{-93}$ & $5367 \pm 40$ & $5320 \pm 65$ & 4.51 & 0.14 & 0.1 & $0.17 \pm 0.01$ & $-4.81 \pm 0.04$ \\
HD 50692 & G0V & 0.7631 & $6060^{+235}_{-118}$ & $5919 \pm 40$ & $5930 \pm 50$ & 4.46 & -0.16 & 0.1 & $0.16 \pm 0.01$ & $-4.82 \pm 0.05$ \\
HD 42250 & G9V & 0.9487 & $5407^{+85}_{-46}$ & $5344 \pm 40$ & $5400 \pm 45$ & 4.48 & -0.05 & 0.1 & $0.16 \pm 0.01$ & $-4.88 \pm 0.04$ \\
HD 38230 & K0V & 1.0214 & $5253^{+61}_{-89}$ & $5077 \pm 40$ & $5130 \pm 70$ & 4.39 & -0.16 & 0.1 & $0.16 \pm 0.01$ & $-4.89 \pm 0.04$ \\
51 Peg & G2IV & 0.817 & $5830^{+214}_{-55}$ & $5817 \pm 40$ & $5800 \pm 115$ & 4.42 & 0.24 & 1.8 & $0.15 \pm 0.01$ & $-4.90 \pm 0.05$ \\
HD 34411 & G1.5IV-VFe-1 & 0.785 & $5979^{+178}_{-140}$ & $5916 \pm 40$ & $5920 \pm 50$ & 4.33 & 0.12 & 0.1 & $0.15 \pm 0.01$ & $-4.90 \pm 0.06$ \\
HD 86728 & G3VaHdel1 & 0.8329 & $5832^{+251}_{-207}$ & $5816 \pm 40$ & $5790 \pm 50$ & 4.45 & 0.25 & 0.2 & $0.15 \pm 0.01$ & $-4.93 \pm 0.06$ \\
HD 145675 & K0V & 1.0049 & $5282^{+251}_{-73}$ & $5300 \pm 40$ &  & 4.46 & 0.37 & 0.1 & $0.16 \pm 0.01$ & $-4.94 \pm 0.04$ \\
HD 210277 & G8V & 0.918 & $5535^{+42}_{-48}$ & $5524 \pm 40$ & $5490 \pm 30$ & 4.47 & 0.22 & 0.3 & $0.15 \pm 0.01$ & $-4.94 \pm 0.05$ \\
HD 221639 & K1V & 1.0866 & $5102^{+139}_{-106}$ & $5091 \pm 40$ & $4900 \pm 40$ & 3.75 & 0.14 & 0.1 & $0.15 \pm 0.01$ & $-4.96 \pm 0.04$ \\
HD 10697 & G3Va & 0.8701 & $5654^{+36}_{-56}$ & $5666 \pm 40$ & $5480 \pm 35$ & 4.13 & 0.14 & 1.1 & $0.14 \pm 0.01$ & $-5.01 \pm 0.06$ \\
HD 145742 & K0 & 1.1996 & $4851^{+13}_{-75}$ & $4801 \pm 100$ &  & 2.85 & -0.05 & 3.6 & $0.15 \pm 0.01$ & $-5.09 \pm 0.06$ \\
HD 89744 & F7V & 0.6919 & $6239^{+23}_{-84}$ & $6254 \pm 40$ & $6070 \pm 110$ & 4.04 & 0.26 & 9.3 & $0.14 \pm 0.01$ & $-5.11 \pm 0.10$ \\
HD 6497 & K2III-IV & 1.3428 & $4514^{+82}_{-71}$ & $4568 \pm 100$ & $4510 \pm 80$ & 2.92 & 0.02 & 3.0 & $0.11 \pm 0.01$ & $-5.41 \pm 0.04$ \\
Gl 705 & K2 & 1.4388 & $4465^{+102}_{-66}$ & $4377 \pm 100$ &  & 2.85 & 0.06 & 3.5 & $0.12 \pm 0.01$ & $-5.44 \pm 0.04$ \\
HD 5857 & G5 & 1.2549 & $4712^{+67}_{-71}$ & $4689 \pm 100$ & $4820 \pm 30$ & 2.92 & -0.2 & 3.1 & $0.11 \pm 0.01$ & $-5.53 \pm 0.21$ \\
\enddata
\end{deluxetable*}
\end{longrotatetable}

\subsubsection{A Note on GJ 4099}
\citet{GJ4099binary} identified a companion to GJ 4099 with 
high-angular-resolution optical imaging
and measured a contrast of $\Delta i = 0.27$ mag (Sloan),
a proper motion indicating that the stars are co-moving, 
and an angular separation of 300-400 mas, 
which \textit{Gaia} DR2 would not resolve \citep[see Figure 1 from][]{Ziegler2018}.

Blended photometry will bias the photometric estimation of \teff. 
Instead, we model the photometry as a binary system. 
First, we assembled the K and M dwarf sample from 
\citet{Mann2015}, which provides \teff\ and metallicity measurements derived from optical and NIR spectroscopy 
as well as synthetic $J, H, K_S$ magnitudes computed 
by integrating the absolutely flux-calibrated optical spectra over each filter. 
Next, we cross-matched this catalog with DR2, 
which provided parallax and the DR2 photometric magnitudes. 
To model the binary, we randomly drew primary and secondary stars with replacement from the sample, 
combined the absolute magnitudes, and assembled the 6-band SEDs
(\textit{Gaia} DR2 $G, G_{\rm BP}, G_{\rm RP}$, 
2MASS $J, H, K_S$),
then minimized $\chi^2$ between the empirical binary SEDs and the observed SED for GJ 4099. 
This procedure yielded temperatures for the primary and secondary of 
\teff\ = 3743 K and 3668 K, respectively. 
HD 209290 has a DR2 \teff\ similar to GJ 4099---we performed a similar fit assuming it is single and found \teff\ = 3875 K (only 7 K cooler 
than its DR2 value).
Whereas DR2 listed this star as only 36 K warmer than GJ 4099, 
we find it to be 132 K warmer than the primary and 207 K warmer than the secondary.
This is consistent with the SME analysis that found 
\teff\ = 3876 K for HD 209290, 
\teff\ = 3682 K for GJ 4099, 
and a difference of 194 K.
Our photometric analysis of GJ 4099 indicates that the primary is 75 K warmer than the secondary, 
which is consistent with the $\Delta i$ measurement from \citet{GJ4099binary}.
Given this small difference in \teff, and the fact that the components 
are well-separated and non-interacting, 
we will assume that our spectrum for GJ 4099 is representative of 
a quiescent 3700 K photosphere.

\section{T\lowercase{i}O Molecular Band Modeling} \label{sec:tio}

\begin{figure*}
    \centering
    \includegraphics[scale=0.75]{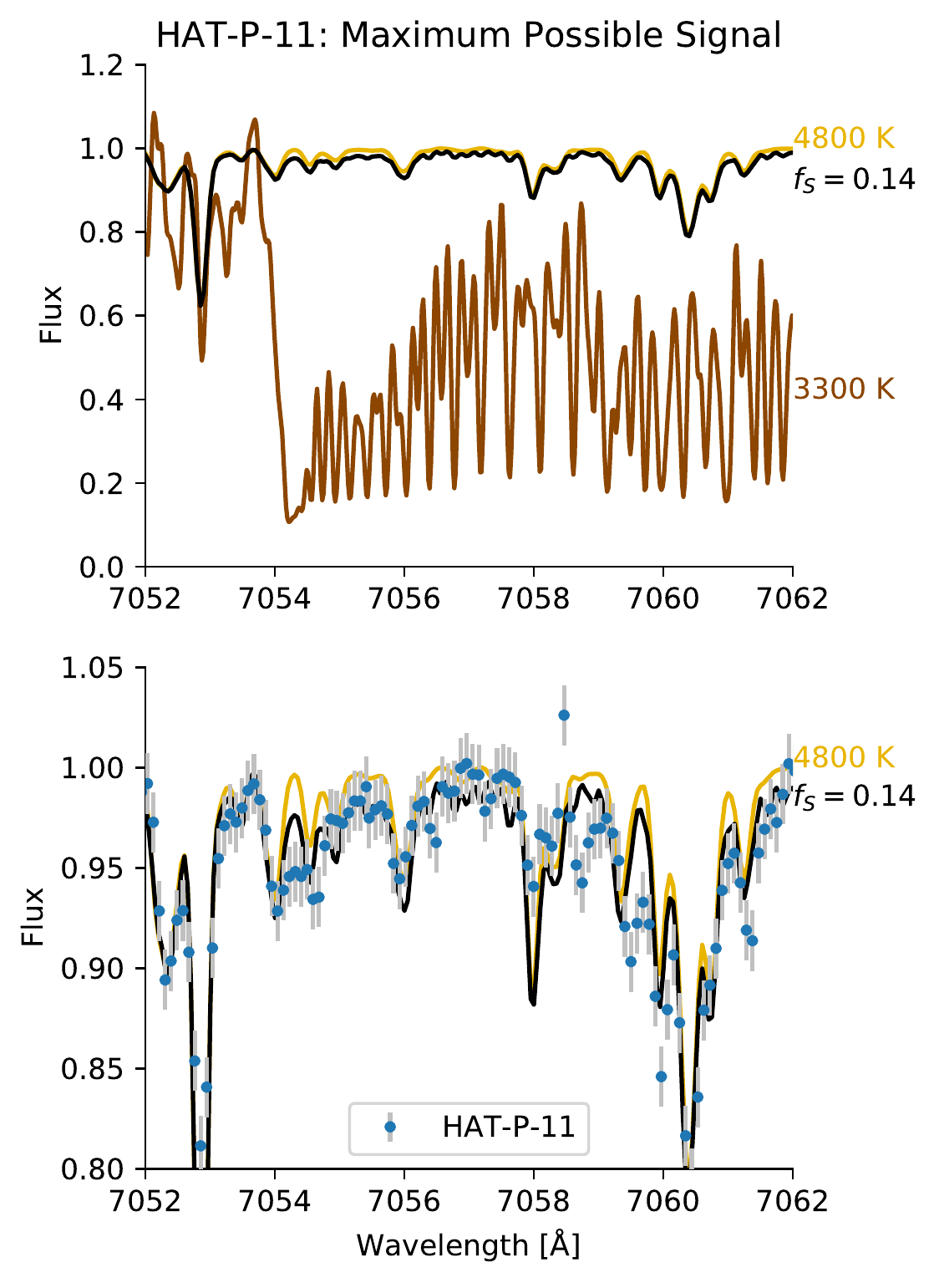}
    \includegraphics[scale=0.75]{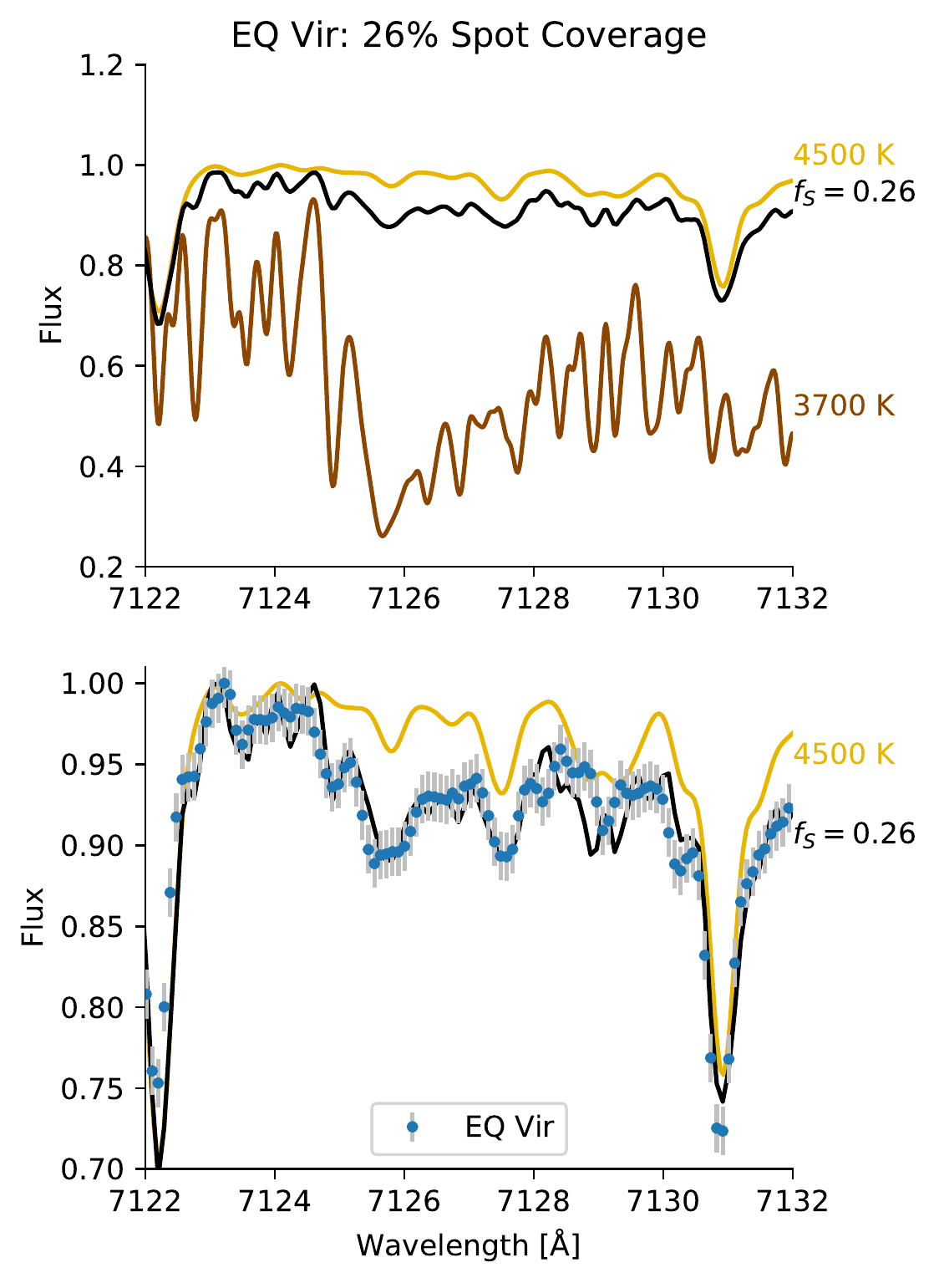}
    \caption{{\it Upper left}: Unspotted PHOENIX model spectrum with the effective temperature of HAT-P-11 (gold) compared with the model spectrum of a cooler star with $\Delta T = 1500$ K (brown). We take the flux-weighted linear combination of the hot and cool components with 14\% spot coverage -- the maximum coverage observed in \citet{Morris2018d} -- to produce the expected spectrum of the spotted star HAT-P-11 (black). Here we assume the most optimistic (and unrealistic) scenario that spots are entirely composed of spot umbrae with temperature difference from the photosphere $\Delta T = 1500$ K for illustrative purposes. In this work, we try to measure the small flux deficit near the TiO bandhead at 7054 \AA\ in the black curve. 
    {\it Lower left}: Zoom in on the difference between the unspotted and spotted composite spectrum (gold and black, respectively), compared with the observed echelle spectrum of HAT-P-11 (blue points). The scatter in the fluxes measured at each wavelength is comparable to the difference between the unspotted and spotted composite models, suggesting that the technique may only marginally detect spot coverage, if any, even in this extreme scenario. 
    {\it Right column}: Same plot for EQ Vir over a different TiO bandhead centered on 7125 \AA, with $f_S=0.26$ and spots with $\Delta T = 800$ K, showing clearer evidence for spot coverage. }
    \label{fig:demo}
\end{figure*}

The spectrum of a spotted star can be approximated as the linear combination of the spectrum of an unspotted star with the effective temperature of the stellar photosphere, and the spectrum of a cooler star with the effective temperature of the starspots, see Figure~\ref{fig:demo}. In principle, there is also a hot component from plages, but the small temperature contrast of plages compared to the mean photosphere makes it difficult to detect in practice.

We select four segments of stellar spectra centered on TiO absorption features for their strong absorption band intensities computed by \citet{Valenti1998}, at 7054, 7087, 7125, and 8859 \AA. 
There may be a small component of TiO absorption in the spectrum of a spotted star like EQ Vir when compared with the inactive star Gl 705, for example, which is a star of similar effective temperature but less chromospheric activity. The goal of this work is to model the spectrum of stars like EQ Vir as a linear combination of the other two spectra (red curve)---one spectrum of a star without TiO absorption, and another with some TiO absorption---in order to infer the spot covering fraction in the observed spectrum (black curve).

The total flux as a function of wavelength, $F_{\rm total}$, that we observe for a given star is a combination of the flux from the photosphere and spot components. In practice, each spectrum is normalized such that the pseudocontinuum is unity, so the linear combination of (hot) photospheric and (cool) spot components is defined by Equation 1 of \citet{Neff1995}:
\begin{equation}
F _ { \text { total } } = \frac { f _ { S } R _ { \lambda } F _ { S } + \left( 1 - f _ { S } \right) F _ { Q } } { f _ { S } R _ { \lambda } + \left( 1 - f _ { S } \right) } \label{eqn:main}
\end{equation}
where $f_S$ is the spot filling factor, $F_Q$ is the spectrum representing the quiescent stellar surface (the photosphere), $F_S$ is the spectrum representing spots, and $R_\lambda$ is the ratio of the spotted to unspotted continuum blackbody flux at a given wavelength. It should be noted that $f_S$ is the fraction of the observable hemisphere of the star that is covered in spots, and may not represent the true total spot coverage of the star, for example, if we are observing a biased hemisphere.

\subsection{Practical application}

We use Markov Chain Monte Carlo
\citep[emcee;][]{Foreman-Mackey2013}
to vary the spot area coverage for each of four TiO absorption regions. 
For a 2 \angstrom-wide window centered on each TiO feature, we fit for a linear combination of the hot and cool component atmospheres. At each step in the chain we regress the two-component model spectrum against the observed stellar spectrum as a function of wavelength to remove any trends that may persist after continuum normalization. 

The uncertainties on the flux within each spectral channel are difficult to compute \textit{a priori}. Thus we marginalize over the flux uncertainty for all spectral channels within the fitting window at each step in the Markov chains, with the appropriate prior penalty on the uncertainties to keep the uncertainties small. 

\subsubsection{Color Prior} \label{sec:colorprior}

We apply a prior to penalize very large spot filling factors, by enforcing that the filling factor is never large enough to skew the broadband colors of the star significantly towards the red. For each star, we compute a Gaussian prior penalty $p$ that a star has an observed \gaia DR2 ($G_\mathrm{BP}-G_\mathrm{RP}$) color $c_O$, compared to the computed two-component mixture model color $c_C$, 
\begin{equation}
\ln p \propto -\frac{1}{2}\left( \frac{c_O - c_C}{\sigma_c} \right)^2,
\end{equation}
where $\sigma_c = 0.006$ is a typical uncertainty on ($G_\mathrm{BP}-G_\mathrm{RP}$) for stars in our sample. Since the signal we are searching for is similar in amplitude to the uncertainties of the observations, fits without this prior tend to prefer non-zero spot filling factors simply because the noisy data can be better fit with two templates than one. The color prior has the important effect of ensuring that non-zero spot filling factors yield a sufficient improvement in the model likelihood to offset the penalty incurred by the color prior.

Spot covering fractions on Sun-like stars are typically small, but span orders of magnitude. In \citet{Morris2017a} we show that the Sun has typical spot coverage near 0.03\%, while the K4V star HAT-P-11 has a typical spot coverage near 3\%.
It has been speculated that the wide range in activity 
could be due in part to the presence of the hot-Jupiter planet in the HAT-P-11 system \citep{Morris2017b}.

\subsubsection{Temperature contrast}

Several efforts exist in the literature to search for correlations between the effective temperatures of stars and the temperatures of their spots \citep[e.g.][]{Berdyugina2005, Mancini2014}. The data sets and analyses are highly heterogeneous and apparent correlations often have several exceptions. Therefore, in order to simplify the analysis in this work, we choose plausible combinations of stellar atmospheres, informed primarily by the temperature contrast of sunspots. 

Sunspots have detailed substructure dominated by the cool penumbra, and the much cooler umbra. Since the ratio of penumbral to umbral area of sunspots is roughly a factor of four, and since the intensity of the penumbra is more than a factor of two greater than the intensity of the umbra in sunspots \citep{Solanki2003}, in this work we will approximate starspots as being entirely composed of penumbra. On the Sun, penumbra typically have temperatures 250-400 K cooler than the mean photosphere. In most cases, we choose combinations of stellar spectra with effective temperature differences of a few hundred Kelvin, manually selecting pairs seeking to minimize the differences in metallicity and surface gravity between the template stars and the target star. Therefore one should not interpret the temperature differences listed in Table~\ref{tab:comptable} as detections; they were adopted \textit{a priori} in order to measure the spot covering fraction under the assumption that starspots on G and K stars generally act like sunspots. The temperatures listed in Table~\ref{tab:comptable} are taken from the SME analysis in Section~\ref{sec:sme}.

\subsubsection{Control experiments}

\startlongtable
\begin{deluxetable*}{ll|ll|ll|rcc}
\tablecaption{Each target star, and the hotter ($T_1$) and cooler ($T_2$) components used to model its spectrum, with spot covering fractions $f_S$ and model comparison with the null hypothesis $\Delta \chi^2$. 
\label{tab:comptable}}
\tablehead{\colhead{Target} & \colhead{$T_\mathrm{eff}$} & \colhead{Comp. 1} & \colhead{$T_1$} & \colhead{Comp. 2} & \colhead{$T_2$} & \colhead{$\Delta T$} & \colhead{$f_S$} & \colhead{$\Delta \chi^2$}}
\startdata
HD 151288 & $4337 \pm 100$ & Gl 705 & $4515 \pm 100$ & HD 209290 & $3876 \pm 100$ & 639  & $0.31 \pm 0.06$ & $197.1$ \\
GJ 702B & $4359 \pm 100$ & Gl 705 & $4515 \pm 100$ & GJ 4099 & $3682 \pm 100$ & 833 & $0.00 \pm 0.06$ &  \\
EQ Vir & $4359 \pm 100$ & Gl 705 & $4515 \pm 100$ & GJ 4099 & $3682 \pm 100$ & 833 & $0.26 \pm 0.06$ & $573.3$ \\
HD 175742 & $4980 \pm 40$ & HD 145742 & $4801 \pm 40$ & Gl 705 & $4515 \pm 100$ & 286  & $0.52 \pm 0.08$ & $21.8$ \\
EK Dra & $5584^{+115}_{-193}$ & HD 210277 & $5524 \pm 40$ & HD 38230 & $5077 \pm 40$ & 447 & $0.31 \pm 0.06$ & $40.1$ \\
Kepler-63 & $5581 \pm 40$ & HD 10697 & $5666 \pm 40$ & HD 221639 & $5091 \pm 40$ & 575 & $0.27 \pm 0.06$ &  \\
HD 45088 & $4786^{+39}_{-36}$ & HD 145742 & $4801 \pm 40$ & Gl 705 & $4515 \pm 100$ & 286 & $0.53 \pm 0.07$ & $21.4$ \\
HD 113827 & $4235 \pm 100$ & Gl 705 & $4515 \pm 100$ & GJ 4099 & $3682 \pm 100$ & 833 & $0.20 \pm 0.06$ & $177.5$ \\
HD 134319 & $5762 \pm 40$ & HD 68017 & $5620 \pm 40$ & HD 145675 & $5300 \pm 40$ & 320  & $0.00 \pm 0.06$ &  \\
HD 127506 & $4651 \pm 100$ & HD 145742 & $4801 \pm 100$ & Gl 705 & $4515 \pm 100$ & 286 & $0.49 \pm 0.08$ & $14.9$ \\
HD 200560 & $5013 \pm 40$ & HD 145742 & $4801 \pm 100$ & Gl 705 & $4515 \pm 100$ & 286 & $0.00 \pm 0.06$ &  \\
HD 220182 & $5363 \pm 40$ & HD 145675 & $5300 \pm 40$ & HD 145742 & $4801 \pm 40$ & 499 & $0.35 \pm 0.06$ & $101.8$ \\
HD 41593 & $5339 \pm 40$ & HD 145675 & $5300 \pm 40$ & HD 145742 & $4801 \pm 40$ & 499 & $0.35 \pm 0.06$ & $102.6$ \\
HD 82106 & $4726 \pm 100$ & HD 145742 & $4801 \pm 100$ & Gl 705 & $4515 \pm 100$ & 286 & $0.00 \pm 0.06$ &  \\
HD 79555 & $4744 \pm 100$ & HD 145742 & $4801 \pm 100$ & Gl 705 & $4515 \pm 100$ & 286 & $0.00 \pm 0.06$ &  \\
HD 87884 & $5047 \pm 40$ & HD 221639 & $5091 \pm 40$ & HD 5857 & $4689 \pm 100$ & 402 & $0.04 \pm 0.10$ &  \\
HD 222107 & $4693 \pm 100$ & HD 6497 & $4568 \pm 40$ & GJ 4099 & $3682 \pm 100$ & 886 & $0.18 \pm 0.06$ & $146.0$ \\
HD 39587 & $6027 \pm 40$ & HD 50692 & $5919 \pm 40$ & HD 62613 & $5489 \pm 40$ & 430 & $0.02 \pm 0.06$ &  \\
HD 149957 & $4389 \pm 100$ & Gl 705 & $4515 \pm 100$ & GJ 4099 & $3682 \pm 100$ & 833 & $0.12 \pm 0.06$ & $60.5$ \\
HD 122120 & $4452 \pm 100$ & Gl 705 & $4515 \pm 100$ & GJ 4099 & $3682 \pm 100$ & 833 & $0.14 \pm 0.06$ & $79.1$ \\
HD 47752 & $4613 \pm 100$ & HD 145742 & $4801 \pm 40$ & Gl 705 & $4515 \pm 40$ & 286  & $0.02 \pm 0.06$ &  \\
HD 148467 & $4253 \pm 100$ & Gl 705 & $4515 \pm 100$ & GJ 4099 & $3682 \pm 100$ & 833 & $0.23 \pm 0.06$ & $245.8$ \\
61 Cyg A & $4336 \pm 100$ & HD 6497 & $4568 \pm 100$ & GJ 4099 & $3682 \pm 100$ & 886 & $0.38 \pm 0.06$ &  \\
HAT-P-11 & $4757^{+154}_{-111}$ & HD 5857 & $4689 \pm 100$ & Gl 705 & $4515 \pm 40$ & 174 & $0.15 \pm 0.06$ &  \\
GJ 9781 A & $4391 \pm 100$ & Gl 705 & $4515 \pm 100$ & GJ 4099 & $3682 \pm 100$ & 833 & $0.21 \pm 0.06$ & $194.8$ \\
HD 110833 & $5004 \pm 40$ & HD 145742 & $4801 \pm 40$ & Gl 705 & $4515 \pm 100$ & 286 & $0.00 \pm 0.06$ &  \\
$\sigma$ Draconis & $5215 \pm 40$ & HD 42250 & $5344 \pm 40$ & GJ 4099 & $3682 \pm 100$ & 1662 & $0.22 \pm 0.06$ & $156.2$ \\
HD 68017 & $5620 \pm 40$ & HD 10697 & $5666 \pm 40$ & HD 221639 & $5091 \pm 40$ & 575 & $0.00 \pm 0.06$ &  \\
HD 67767 & $5264 \pm 40$ & HD 182488 & $5367 \pm 40$ & GJ 4099 & $3682 \pm 100$ & 1685 & $0.15 \pm 0.06$ & $61.7$ \\
HD 73667 & $4800 \pm 40$ & HD 221639 & $5091 \pm 40$ & Gl 705 & $4515 \pm 100$ & 576 & $0.00 \pm 0.06$ &  \\
Kepler-17 & $5913 \pm 40$ & HD 50692 & $5919 \pm 40$ & HD 62613 & $5489 \pm 40$ & 430 & $0.36 \pm 0.06$ &  \\
HD 62613 & $5489 \pm 40$ & HD 10697 & $5666 \pm 40$ & HD 221639 & $5091 \pm 40$ & 575 & $0.16 \pm 0.06$ & $10.0$ \\
HD 14039 & $5072^{+98}_{-73}$ & HD 221639 & $5091 \pm 40$ & HD 5857 & $4689 \pm 100$ & 402 & $0.02 \pm 0.06$ &  \\
51 Peg & $5817 \pm 40$ & HD 86728 & $5816 \pm 40$ & GJ 4099 & $3682 \pm 100$ & 2134 & $0.10 \pm 0.06$ & $24.9$ \\
HD 34411 & $5916 \pm 40$ & HD 86728 & $5816 \pm 40$ & HD 145675 & $5300 \pm 40$ & 516 & $0.00 \pm 0.06$ &  \\
\enddata
\end{deluxetable*}

We construct a simple test to ensure that the two-component model is not over-fitting the spectra, and produces reliable uncertainties. We fit a sample of inactive stars with a linear combination of two stars: the spectrum of the same star and the spectrum of a cooler star. In this control run, we expect that the maximum-likelihood model should choose a spot covering fraction $f_S \approx 0$, since the spectrum of the target star should be a perfect fit to itself. For a sample of 10 control stars, we find typical $f_S < 0.06$ (68\% confidence). We interpret this result to mean that the smallest reliable uncertainty on $f_S$ should be $\sim 0.06$; so throughout this work, the uncertainty on $f_S$ is the maximum of the formal uncertainty or 0.06.

We also analyze spectra of the twilight sky as a proxy for solar spectra, on a day with no sunspots. We model the solar spectrum as a linear combination of HD 10697 ($T_Q = 5654^{+36}_{-56}$ K) and HD 221639 ($T_S = 5102^{+139}_{-106}$ K) and measure $f_S \lesssim 0.001$. This measurement is so small due largely to the very high signal-to-noise of the solar observations in comparison with the stellar observations.

\subsection{Validation on Previously Characterized Stars}

\begin{figure}
    \centering
    \includegraphics[scale=0.8]{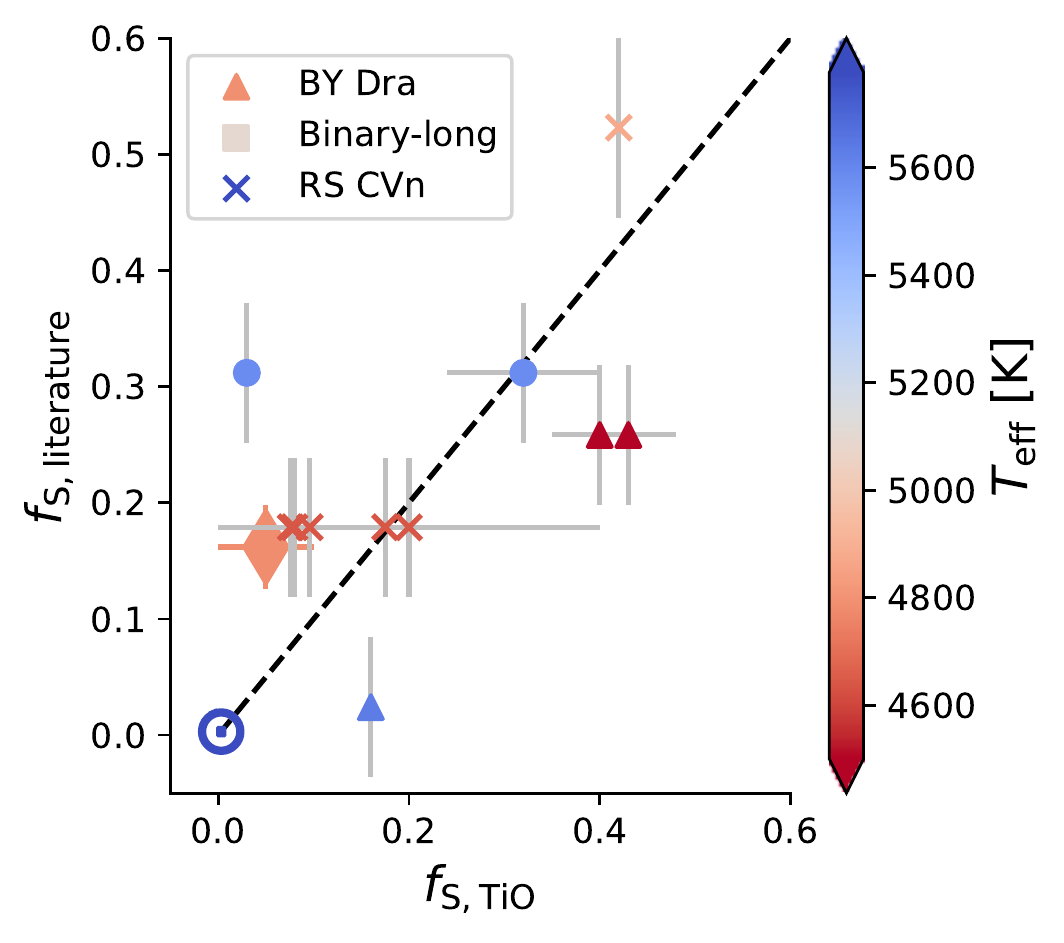}
    \caption{Comparison of literature spot coverage fractions to our results. The Sun is shown with the ``$\odot$'' symbol, HAT-P-11 is shown with the large diamond. The most significant outlier is HD 134319 (IU Dra), for which we measure very small spot covereage (see Section~\ref{sec:iudra}). The scatter about the one-to-one correspondance line represents a combination of: (1) intrinsic variability in the spotted surface area of each star, and (2) underestimated uncertainties in one or both measurements.}
    \label{fig:fscomp}
\end{figure}

In the following subsections we compare our results to previous results in the literature, summarized in Figure~\ref{fig:fscomp}.

\subsubsection{EQ Vir} \label{sec:eqvir}

EQ Virginis is a K5Ve flare star in the nearby cluster IC 2391 
\citep[50 Myr;][]{Barrado2004}
with rotation period $P_\mathrm{rot} = 3.9 \pm 0.1$ d \citep{Montes2001,Paulson2006,Suarez2016}.  \citet{Saar2001} assumed $T_Q = 4380$ K and $T_S = 3550$ K, and found a spot covering fraction $f_S = 0.43 \pm 0.05$. Subsequently, \citet{ONeal2004} 
adopted 61 Cyg A as the quiescent template 
\citep[\teff = 4374 K;][]{Heiter2015}, and 
measured $T_S = 3350 \pm 115$ K with $f_S$ = 0.34-0.45.

We model the spectrum of EQ Vir as a linear combination of the spectra of Gl 705 (K2V, $T_Q = 4465^{+102}_{-66}$ K) and GJ 4099 (M2V, $T_S = 3682$ K), 
which is a warmer spot temperature than the prior two studies. 
We recover spot covering fractions $f_S = 0.25 \pm 0.06$, smaller than the spot covering fractions measured by earlier observers. In large part, the discrepancy is due to our use of a strong prior on the color of the star (see Section~\ref{sec:colorprior}), which places strict upper limits on the plausible spot covering fractions.

To compare our results with \citet{Saar2001} and \citet{ONeal2004}, and to understand the effect of the color prior, 
we compute the expected \gaia DR2 ($G_\mathrm{BP}-G_\mathrm{RP}$) color of EQ Vir, given the spot properties of \citet{ONeal2004}, and compare it to its observed color. Its photospheric temperature measured via SME is $T_\mathrm{eff} = 4519 \pm 200$ K, and synthetic photometry of a PHOENIX model atmosphere with this temperature has color $(G_\mathrm{BP}-G_\mathrm{RP})_Q$=1.4772. \citet{ONeal2004} assume the spotted component has temperature $T_S = 3350 \pm 115$ K, which from synthetic photometry should have color $(G_\mathrm{BP}-G_\mathrm{RP})_S$=2.9739. Combining these colors with spot covering fraction $f_S = 0.43$ using Equation~\ref{eqn:color} yields the expected composite color ($G_\mathrm{BP}-G_\mathrm{RP})=1.8135$. The observed \gaia EQ Vir color ($G_\mathrm{BP}-G_\mathrm{RP})=1.4960 \pm 0.0054$ is 58$\sigma$ discrepant with the \citet{Saar2001} observation, and up to 62$\sigma$ discrepant with the range reported by \citet{ONeal2004}. We conclude that the spot covering fraction must be $f_S < 0.43$ in order not to redden the star significantly beyond its actual color, assuming the effective temperature measured by SME is the correct photospheric temperature.

To verify that the PHOENIX model atmospheres are not responsible for this large discrepancy, we also empirically estimated the DR2 colors by constructing a color--temperature relation based on the \citet{Boyajian2012} K/M benchmark sample, which covers  $3054 < \teff < 5407$ K.\footnote{$(G_\mathrm{BP}-G_\mathrm{RP}) = 11.7\; -3.91\times10^{-3}\; \teff\;  +3.57\times10^{-7}\; \teff^2$,  and is valid over $3054 < \teff < 5407$ K.}
This predicts $(G_\mathrm{BP}-G_\mathrm{RP})_S$=2.608 for 3350 K and 1.324 for 4515 K, and we find a 20$\sigma$ discrepancy with the observed color. This confirms that there is a significant discrepancy between the predicted and observed colors of EQ Vir using the \citet{ONeal2004} spot coverage and temperatures.

For completeness, we carry out the same calculation using our maximum-likelihood model for EQ Vir. We adopt Gl 705 for $T_Q$ with $(G_\mathrm{BP}-G_\mathrm{RP})_Q$ = 1.4388, and GJ 4099 for $T_S$ with $(G_\mathrm{BP}-G_\mathrm{RP})_S$ = 2.1969. Our maximum-likelihood spot covering fraction is $f_S = 0.24$, which gives a predicted color $G_\mathrm{BP}-G_\mathrm{RP} = 1.4976$, which is $<1\sigma$ consistent with the observed color of EQ Vir. 

\subsubsection{EK Dra (HD 129333)}

EK Draconis is a G1.5V BY Dra variable generally thought to resemble the young Sun \citep{Montes2001}. \citet{Jarvinen2007} used atomic line profiles to detect both high and low-latitude spots with $\Delta T = 500$ K. Zeeman Doppler imaging from \citet{Waite2017} reveals $P_\mathrm{rot} = 2.766 \pm 0.002$ d, and an amazing variety of spots---sometimes at mid-latitudes, sometimes at the pole, covering 2-4\% of the stellar surface. \citet{ONeal2004} measured a much larger spot coverage $f_S = $0.25-0.40 for $T_Q = 5830$ K and $T_S \gtrsim 3800$ K. 

We model the spectrum of EK Dra with a linear combination of spectra from HD 210277 ($T_Q = 5535^{+42}_{-48}$ K) and HD 38230 ($T_S = 5253^{+61}_{-89}$ K). The only TiO band that yields informative constraints on the spot covering fraction is at 7054 \angstrom, predicting a spot covering fraction of $f_S = 0.28 \pm 0.06$, consistent with the lower end of the coverage measured by \citet{ONeal2004}. The higher spot covering fractions are excluded by the color prior (see previous section for a detailed discussion).

\subsubsection{HAT-P-11}

HAT-P-11 is an active K4V dwarf in the \kepler field with a transiting hot Jupiter. Transits of the stellar surface reveal frequent starspot occultations \citep{Deming2011,Sanchis-Ojeda2011}, which yield spot covering fractions $f_S =$ 0.0-0.1 within the transit chord; while rotational modulation predicts $f_S \gtrsim 0.03^{+0.06}_{-0.01}$ \citep{Morris2017a}. HAT-P-11 appears to have a $\sim$10 year activity cycle, and may be modestly more chromospherically active than planet hosts of similar rotation periods \citep{Morris2018b}. Recent ground-based photometry of spot occultations within the transit chord yielded spot coverage $f_S = 0.14$ \citep{Morris2018d}. 

\begin{figure}
    \centering
    \includegraphics[scale=0.8]{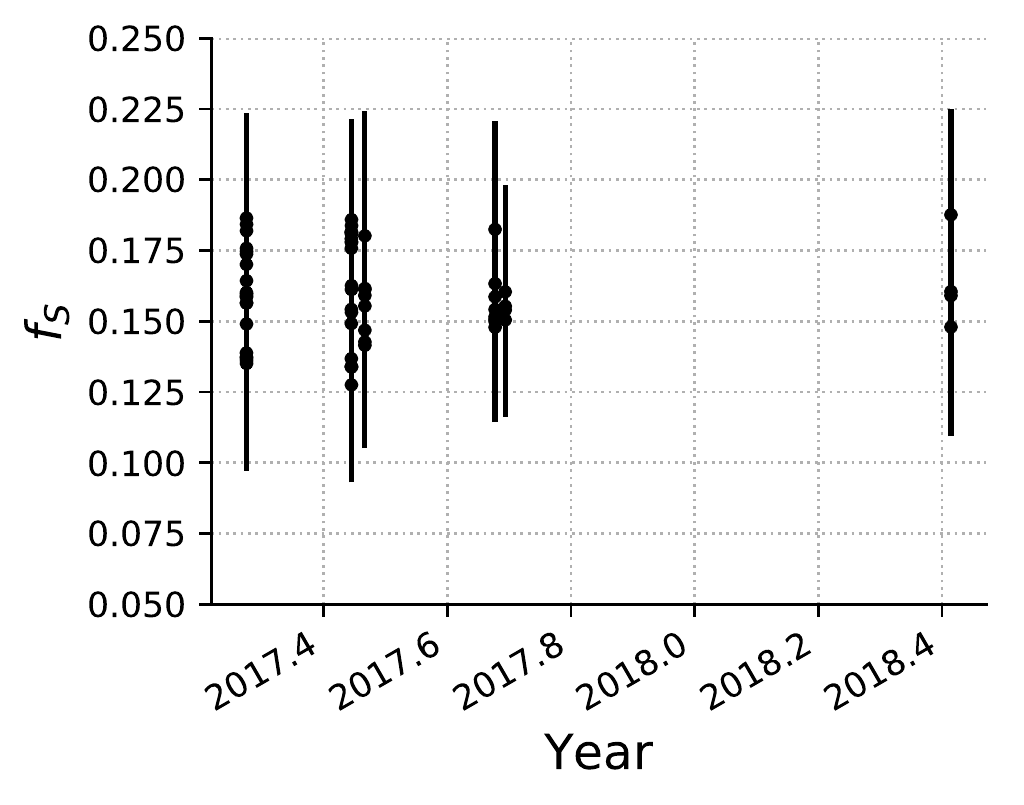}
    \caption{The spot covering fraction $f_S$ of HAT-P-11 inferred via TiO absorption from starspots in the spectrum of HAT-P-11 as a function of time. }
    \label{fig:h11}
\end{figure}

It is a valuable validation exercise to apply the TiO molecular band modeling technique to HAT-P-11 and to compare with the results of spot occultation analysis. We model the spectrum of HAT-P-11 as a linear combination of the spectra of HD 5857 ($T_Q = 4712^{+67}_{-71}$ K) and Gl 705 ($T_S = 4465^{+102}_{-66}$ K), giving the spots $\Delta T_\mathrm{eff} = 250$ K, similar to typical sunspot penumbra. We find a consistent spot covering fraction between observations separated by a year near $f_S \approx 0.15 \pm 0.06$, see Figure~\ref{fig:h11}. This is 
broadly consistent with the spot covering fractions measured in transit by \citet{Morris2017a}, if on the high end, and consistent with the more recent observations of \citet{Morris2018d}. The TiO estimate might be slightly larger than the typical transit chord crossing spot filling factor because of the small temperature difference that we have assumed here (250 K)---a larger temperature difference would produce a correspondingly smaller spot covering fraction. Perhaps this is evidence that the correct spot temperature should be a bit cooler than our assumed $T_S$. 

\subsubsection{HD 222107 ($\lambda$ And)} \label{sec:lamand}

$\lambda$ Andromedae is a single-lined spectroscopic binary (SB1) with an old disk giant primary identified as an RS CVn variable with rotation period $P_\mathrm{rot} = 54$ d, which has at least two large starspots \citep{Calder1938, Landis1978, Bopp1980, Poe1985, Padmakar1999, Drake2011, Pandey2012, Parks2014}. $BV$ photometry of the rotational modulation predicts one large spot covering 8\% of the stellar surface, with an estimated temperature $T_S = 4000 \pm 300$ K \citep{Donati1995}. \citet{oneal1998} modelled TiO molecular band absorption ranging $f_S =$ 0.14-0.21. \citet{Mirtorabi2003} developed a TiO photometric index to measure the spot covering fraction over several years and found spot coverage varying from a few to $\lesssim 40\%$. \citet{Frasca2008} measured simultaneous photometry and spectroscopy of $\lambda$ And and found $f_S = 0.076^{+0.023}_{-0.014}$. Using the CHARA array, \citet{Parks2014} found $f_S = 0.096$.

We model $\lambda$ And with a combination of HD 6497 ($T_Q = 4514^{+82}_{-71}$ K) and GJ 4099 ($T_S = 3846^{+181}_{-190}$ K). We find $f_S = 0.18 \pm 0.06$, consistent with \citet{Donati1995, oneal1998, Donati2003, Frasca2008, Parks2014}. 

Due to the RS CVn nature of $\lambda$ And, it appears as an outlier in Figure~\ref{fig:fs_rotation}, having a very long rotation period but significant, nonzero spot coverage. The apparently high spot coverage at its slow rotation period may be an effect of the low surface gravity of this sub-giant primary star. The low \logg\ of $\lambda$ And implies a longer convective timescale $\tau_c$, and therefore the Rossby number of the star $Ro = P_{\rm rot} / \tau_c$ may still be small even though its rotation period is long.

\subsubsection{HD 45088 (OU Gem)}
This is a double-lined spectroscopic binary (SB2), with an orbital period 
$P_\mathrm{orb} = 6.99$ d \citep{Griffin1975, Bopp1980}.
\citet{Glazunova2014} solved its orbit, characterized the components, and found the following properties 
for the primary and secondary, respectively:
\teff\ = 5025 and 4508 K,
\logg\ = 4.3 and 4.5 dex, 
and [Fe/H] $= -0.2$ dex.

\citet{ONeal2001} accounted for the secondary in their analysis and fixed the primary and secondary temperatures to 
4925 K and 4550 K, then fit for a third component assumed to have a temperature of 
3525 K and a spectrum from Gl 96, 
and found $f_S = 0.20-0.35$. 
\citet{Glazunova2014} also noted temporal changes in chemical abundances consistent with substantial spot coverage. 

We modeled the TiO features of HD 45088 as a combination of HD 145742 ($T_Q=4851^{+13}_{-75}$ K) and Gl 705 ($T_S = 4465^{+102}_{-66}$ K), 
ignoring the binarity, found apparent ``spot covering fraction'' $f_S = 0.52 \pm 0.14$, which in this case can be instead interpreted as the fractional flux contributed by the secondary component.
According to a 3 Gyr PARSEC isochrone model with [Fe/H] = $-0.2$ dex \citep{Bressan2012}, 
the stellar radii at our component temperatures are 
$R_A = 0.69$ $R_\odot$ and 
$R_B = 0.63$ $R_\odot$;
the ratio of the cooler area from the secondary to the total area 
of both stars is then 0.48, which 
is consistent with the spot covering fraction we inferred. 
The difference between \citet{ONeal2001} and our procedure is that we did not account for the secondary's flux, but instead 
fit its photosphere and returned the relative stellar areas instead of the spot covering fraction. This exercise validates our TiO fitting procedure, 
at least in this ``heavily spotted'' regime,
as the spot area fractions are consistent with the stellar flux contributions that are expected for this stellar binary.

\subsubsection{HD 175742 (V775 Her)}

HD 175742 (K0V) is also an RS CVn variable, with rotation period $P_\mathrm{rot} = 2.88$ d \citep{Rutten1987}. \citet{Alekseev2000} 
estimated the spectral type of the companion to be M3V 
($\Delta T \approx 1900$ K), 
then measured the spot covering fraction via rotational modulation and estimated $f_S \lesssim 0.42$ with $\Delta T = 900$ K. 

We model HD 175742 with a combination of HD 145742 ($T_Q = 4851^{+13}_{-75}$ K) and Gl 705 ($T_S = 4465^{+102}_{-66}$ K) and find $f_S = 0.54 \pm 0.13$, consistent with \citet{Alekseev2000}.

\subsubsection{Kepler-17}

\citet{Desert2011, Bonomo2012, davenportthesis} measure positions and differential rotation of spots for this star with $P_\mathrm{rot} = 12.2$ d. \citet{Estrela2016} measure activity cycle $P_\mathrm{cyc} = 1.12 \pm 0.16$ year. We measure $f_S = 0.16 \pm 0.06$, qualitatively consistent with the relatively large spot covering fraction necessary to produce spot occultations in nearly every transit, though we could not find a precise spot coverage number in the literature.

\subsubsection{HD 134319 (IU Dra)} \label{sec:iudra}

\citet{Messina1998} use rotational modulation to predict a spot covering fraction of at least $f_S \gtrsim 0.16$ for this BY Dra-type variable. Adaptive optics imaging revealed that HD 134319 has a M4.5V companion \citep{Mugrauer2004}, but its late type and large separation from the host star of (5\farcs5) makes it unlikely that it is a significant source of contaminating flux. We model HD 134319 with a combination of HD 210277 ($T_Q = 5535^{+42}_{-48}$ K) and HD 38230 ($T_S = 5253^{+61}_{-89}$ K), and find $f_S = 0.0 \pm 0.06$.

\section{Discussion} \label{sec:discuss}

\subsection{Major caveats}

\begin{figure}
    \centering
    \includegraphics[scale=0.75]{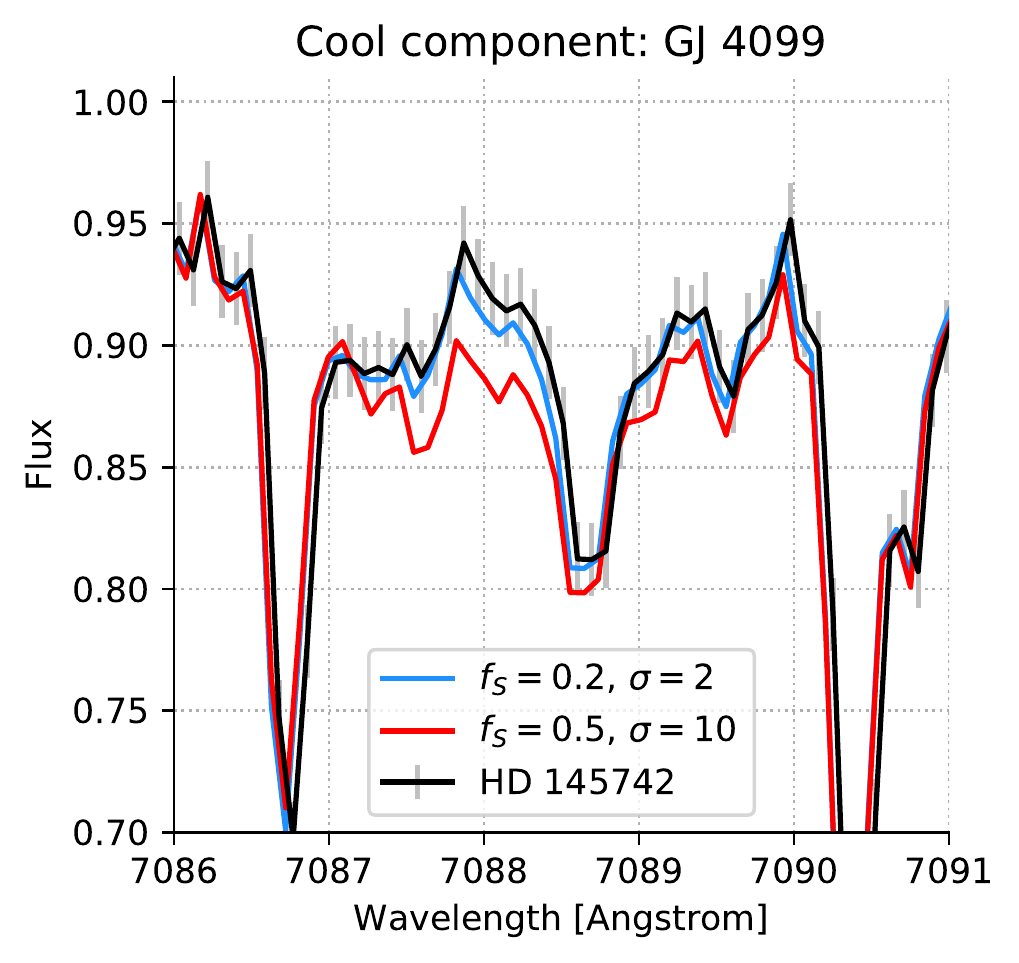}
    \caption{Synthetic two-component spectra created by combining HD 145742 with various spot coverages represented by the spectrum of GJ 4099. The flux-weighting of the spots conspires to make it extremely difficult to measure small spot coverage, though larger spot coverages can be measured robustly.}
    \label{fig:demo2}
\end{figure}

The spot coverage that we infer from this TiO modeling technique is explicitly a function of the spot temperature that we assume. In this work, we have simply adopted a spot temperature contrast for each star, typically of a few hundred Kelvin, which neglects to explore the degeneracy between spot coverage and spot temperature. Our spot coverage uncertainties are therefore necessarily underestimated by not accounting for the uncertainty in the spot temperatures. 

Another point of clarification must be made about from where we expect TiO absorption to originate, which is different for G and K stars. TiO forms at temperatures below $T \lesssim 4000$ K. On the Sun, umbrae are 3900-4800 K and penumbrae are 5400-5500 K \citep{Solanki2003}. Therefore in a Sun-like star, we should expect TiO absorption to only be generated by the umbrae of starspots. However, if we assume that starspots have a similar temperature contrast on a mid-K star with 4400 K, penumbrae may have temperatures near 4000 K, implying that the entirety of starspots, umbrae and penumbrae together, generate TiO absorption. As a result, the spot coverages for stars earlier than $\sim$K5 should be interpreted as umbral spot coverages, while spot coverages for stars later than $\sim$K5 may be interpreted as penumbral-plus-umbral spot coverages. 

The challenge of making TiO absorption from $R\approx31,500$ spectroscopy is illustrated in Figure~\ref{fig:demo2}, where we have constructed two artificial spotted spectra from HD 145742 by adding cooler components with the spectrum of GJ 4099 via Equation~\ref{eqn:main}. Spot coverages as large as 20\% make insufficient differences from the original unspotted spectrum to robustly claim a spot coverage, and it is only in the $\sim 50\%$ coverage regime that truly significant spot coverages make obvious changes in the observed spectrum of the spotted star. It is also clear how sensitive the technique may be to continuum normalization, as any misinterpreted curvature in the continuum could easily over- or under-estimate the spot coverage in TiO bandheads. Using ARCES spectra and the technique presented here, we caution the reader that our spectra are sensitive enough to robustly identify spot coverage $>30\%$ ($5\sigma$) or more, and give suggestive results for small spot coverages which are primarily constrained by the \gaia color prior rather than the echelle spectroscopy. 

\subsection{\logrprime as a proxy for spottedness}

\begin{figure}
    \centering
    \includegraphics[scale=0.8]{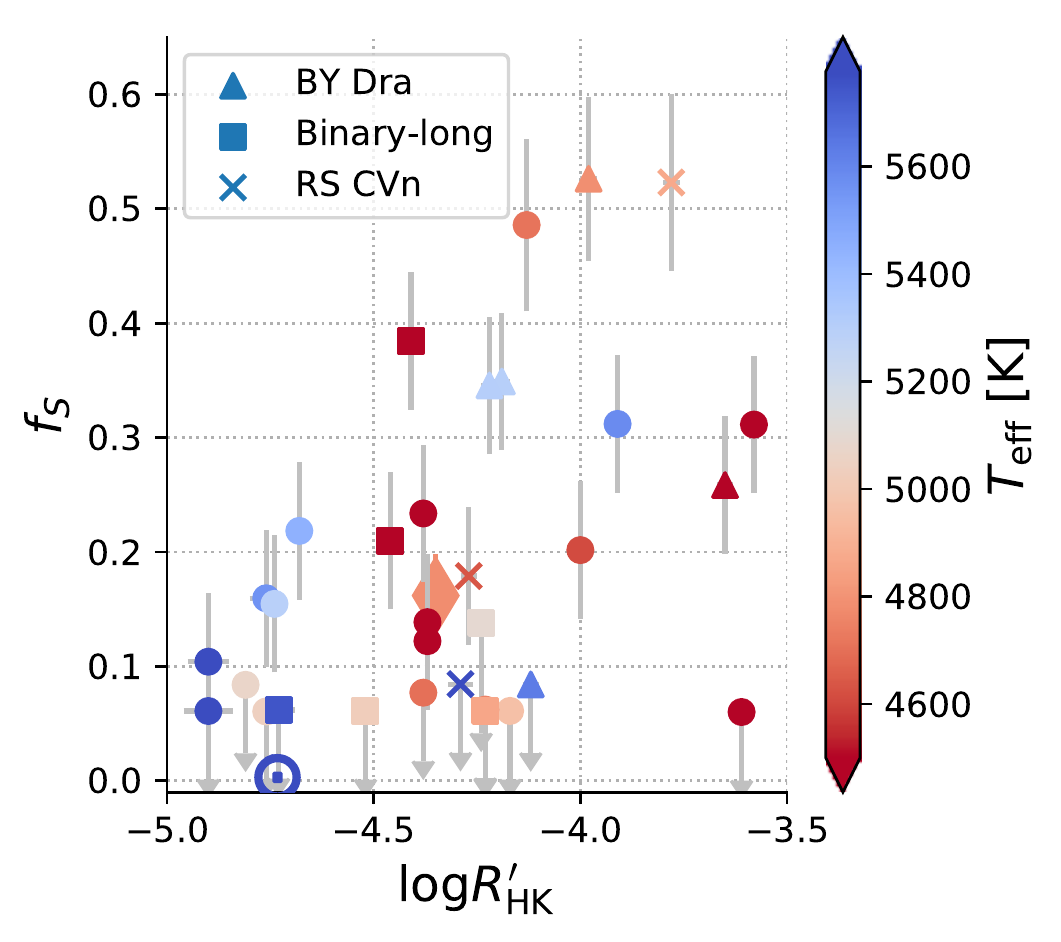}
    \caption{Spot covering fraction $f_S$ as a function of \logrprime. The ``$\odot$'' is the Sun, the large rhombus is HAT-P-11. Circular points mark stars that have no special classification. }
    \label{fig:fs_rprime}
\end{figure}

We compare the spot covering fractions $f_S$ with the chromospheric activity index \logrprime in Figure~\ref{fig:fs_rprime}. We use the \logrprime index rather than the $S$-index to account for the different baseline photospheric emission in stellar atmospheres as a function of effective temperature; and indeed the results appear to be independent of stellar $T_\mathrm{eff}$. There is a dearth of stars with high spot covering fractions and low \logrprime, in other words, stars with relatively inactive chromospheres stars tend to have relatively small spot coverages. As chromospheric activity increases, the maximum observed spot coverage increases, with stars distributed throughout a broad range of spot coverages at each chromospheric activity level. 

We note the apparent discrepancy between the measured spot coverages for HAT-P-11 measured via starspot occultations \citep{Morris2017a} and TiO absorption (orange diamond), for which there are several explanations. The spot coverage measured in transit via spot occultations varied between $0 < f_S < 0.1$ from transit to transit between 2009-2013. The echelle spectra presented here were collected from 2017-2018, at a different phase in the stellar activity cycle \citep{Morris2017b}, where the apparent spot coverage measured in transit is somewhat higher, $f_S \sim 0.14$ \citep{Morris2018d}. The later spot coverage estimate is consistent with the spot coverage measured from molecular band modeling, $f_S \sim 0.12 \pm 0.06$. 

Several of the spot coverage measurements are consistent with $f_S \gtrsim 0.5$, which begs the question: are we measuring spot coverage or evidence for a different mean photospheric temperature (i.e. the Zebra effect, \citealt{Pettersen1992})? Given recent detailed observations by \citet{Gully2017} which yield $f_S \sim 0.8$, we follow their example and simply state the coverage $f_S$ of the apparently cooler surface, noting that in the large coverage regime ($f_S \gtrsim 0.5$), it might be perilous to think of $f_S$ as ``starspot coverage'' and instead it would be more accurate to think of it as the ``cool atmosphere component filling factor.'' It is also possible that an unseen late-type companion could be contributing to the apparently large spot covering fraction. 

The most active stars are HD 175742 ($f_S = 0.54 \pm 0.13$), HD 45088 ($f_S = 0.54 \pm 0.13$) and Gl 702 B ($f_S = 0.45 \pm 0.06$), 
two of which are short-period spectroscopic binaries.
These three K stars are also some of the fastest rotators (Table ~\ref{tab:rotation}). 

We note that none of the stars studied here are candidates for Maunder minima stars \citep{Wright2004b}. The template spectra of stars with \logrprime $\lesssim -5.1$ are consistent with having evolved off the main sequence, judging by measured surface gravities and their positions on the color magnitude diagram in Figure~\ref{fig:cmd}. 

\subsection{Spottedness as a function of rotation}

\begin{figure}
    \centering
    \includegraphics[scale=0.8]{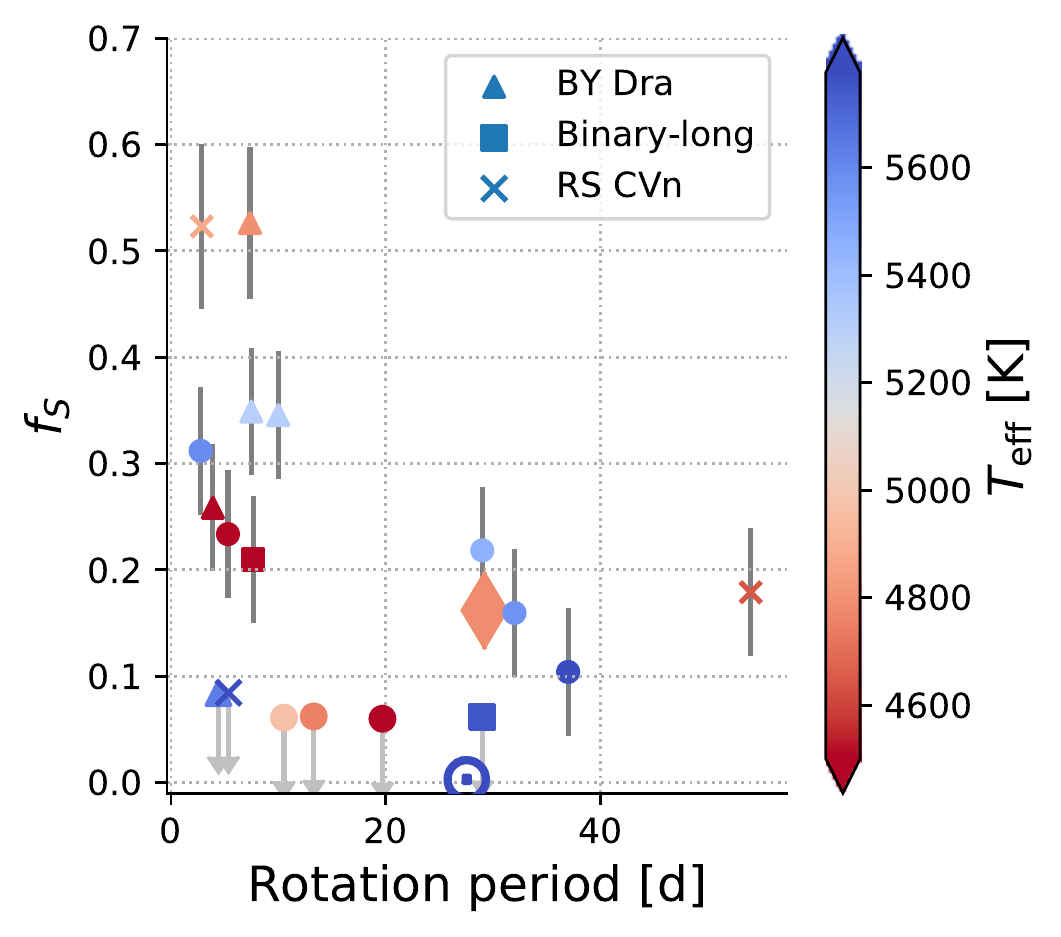}
    \caption{Spot covering fraction $f_S$ as a function of \logrprime. The $\odot$ is the Sun, the large rhombus is HAT-P-11. The outlier at the longest rotation period is the RS CVn star $\lambda$ And (HD 222107), see Section~\ref{sec:lamand} for more details.}
    \label{fig:fs_rotation}
\end{figure}

The spot covering fraction as a function of stellar rotation period is shown in Figure~\ref{fig:fs_rotation}, with rotation periods listed in Table~\ref{tab:rotation}. 
Figure~\ref{fig:fs_rotation} is essentially the reverse of Figure~\ref{fig:fs_rprime}, since rotation period and chromospheric activity are inversely related 
via the activity--Rossby number relation \citep[i.e., faster rotators are more active;][]{Noyes1984}. At short rotation periods there is a broad range of spot coverages, while the coverage generally declines with increasing rotation period.

One exception is an active star with a long period -- the G8IVk star HD 222107 ($\lambda$ And) with $P_\mathrm{rot} = 54$ d \citep{Rutten1987}. This star is discussed in detail in Section~\ref{sec:lamand}, so we simply note here that it may have a non-zero spot coverage despite its long rotation period because it has been classified as an RS CVn variable. 

It is interesting to note that the transition to significantly spotted stars near \logrprime$> -4.3$ roughly corresponds to the transition noted in Rossby number--\logrprime space by \citet{Mamajek2008}. The authors find that stars with \logrprime$> -4.3$ show a weak correlation with Rossby number, while the correlation is much tighter for more negative \logrprime (see Figure 7 of \citealt{Mamajek2008}).

\begin{deluxetable*}{lcr}
\tablecaption{Literature rotation periods. \label{tab:rotation}}
\tablehead{\colhead{Target} & \colhead{$P_\mathrm{rot}$ [d]} & \colhead{Reference}}
\startdata
EK Dra & 2.766 & \citet{Waite2017} \\
HD 175742 & 2.88 & \citet{Rutten1987} \\
EQ Vir & 3.9 & \citet{Suarez2016} \\
HD 134319 & 4.43 & \citet{Wright2011} \\
HD 148467 & 5.31 & \citet{Houdebine2016} \\
HD 39587 & 5.36 & \citet{Wright2011} \\
Kepler-63 & 5.407 & \citet{McQuillan2013} \\
HD 45088 & 7.36 & \citet{Rutten1987} \\
HD 220182 & 7.49 & \citet{Wright2011} \\
GJ 9781A & 7.66 & \citet{Houdebine2016} \\
HD 41593 & 10.0 & \citet{Isaacson2010} \\
HD 200560 & 10.526 & \citet{Strassmeier2000} \\
Kepler-17 & 12.159 & \citet{McQuillan2013} \\
HD 82106 & 13.3 & \citet{Pizzolato2003} \\
GJ 702B & 19.7 & \citet{Wright2011} \\
$\sigma$ Draconis & 29.0 & \citet{Pizzolato2003} \\
HD 68017 & 29.0 & \citet{Isaacson2010} \\
HAT-P-11 & 29.2 & \citet{Morris2017a} \\
HD 62613 & 32.0 & \citet{Isaacson2010} \\
51 Peg & 37.0 & \citet{Wright2011} \\
HD 222107 & 54.0 & \citet{Rutten1987}
\enddata
\end{deluxetable*}

\subsection{The photosphere-chromosphere connection}

Let us now connect these synoptic measurements with the small-scale magnetic activity occurring on the surfaces of these stars. Figure~\ref{fig:fs_rprime} is essentially comparing the chromospheric emission in plages, measured in terms of \logrprime, to the photospheric absorption by starspots, in terms of $f_S$. Plages are magnetically active regions which are bright in stellar chromospheres, leading to emission in the \ion{Ca}{2} H \& K line cores \citep{Hall2008}.  Starspots are regions of intense magnetic fields where convection is suppressed in the photosphere \citep{Solanki2003}. 

On the Sun, we can spatially resolve plages and sunspots, and we see: (1) that plages emerge in the same active latitudes as sunspots throughout the solar activity cycle; (2) plage regions are generally larger than sunspots; and (3)  sunspots are accompanied by plage regions, while the reverse is not always true (plages may emerge without accompanying sunspots) \citep{Mandal2017}.

We propose that the observations in Figure~\ref{fig:fs_rprime} show a related trend in the ensemble of stars. Stars with relatively small \logrprime host plage regions, which generate the observed \ion{Ca}{2} H \& K emission. A transition occurs near $\logrprime \sim -4.3$, where some of the magnetic active regions on the star must now be large enough to produce both plages \textit{and} substantial coverage by starspots, resulting in the observed response in $f_S$ for $\logrprime > -4.3$. 

This may explain why the spot coverage $f_S$ does not increase monotonically with increasing \logrprime, since plages are not always accompanied by starspots, so a range of spot coverages can be consistent with a particular\logrprime. This may also explain the dearth of stars with substantial spot coverage $f_S$ but small \logrprime, since starspots are always accompanied by plages. 

\subsection{BY Draconis variability}

Several of the stars with non-zero spot coverages have already been noted as BY Draconis-type variable stars, indicating that they show rotational modulation generally attributed to asymmetric starspot coverage \citep{Kron1950, Chugainov1966, Jaschek1990}. Some of the stars in the sample have not been identified as BY Dra-type variables despite having significant spot coverage, such as HD 151288 and Gl 702 B, which may indicate one of two scenarios: either time-series photometry has yet to reveal the rotational modulation of these stars, or perhaps the distribution of starspots is axisymmetric, causing minimal rotational variability. 

The latter scenario---where symmetric spot distributions produce little rotational variability but apparently significant TiO absorption---highlights an important feature of molecular band modeling for starspot hunting. Molecular modeling does not need the distribution of spots to be asymmetric in order to detect starspots. 

It is worth asking if the BY Draconis variable designation has meaning in the age of \kepler photometry, which has revealed that innumerable stars are rotationally variable once sufficiently precise photometry is available \citep[e.g.:][]{Walkowicz2013,McQuillan2013,McQuillan2014,Giles2017}. We propose that the BY Dra label be reserved for stars with large rotational variability, for example, greater than 0.01 mag. This might preserve the BY Dra designation for only those targets with unusually large rotational variability, without admitting most main-sequence stars into the class.

\section{Conclusions} \label{sec:conclusion}

Detailed, independent analyses of the effective temperatures of our sample of 55 F, G, K, and M stars using both SME and \moog show good agreement with the \gaia DR2 effective temperature catalog \citep{DR2prop}. The \gaia effective temperatures inferred from broadband colors appear not to be strongly affected by stellar activity or spectral type, reaffirming the importance and accuracy of the full \citet{DR2prop} catalog.

We have used molecular band modeling of four strong TiO absorption features in the optical to measure spot coverage on active G and K stars. We find consistent spot coverages with previous results on several benchmark stars, and provide uniform spot coverage measurements for a sample of 29 active stars spanning a range of activity levels. We find that color priors are critical for constraining the spot coverage on active stars while preserving their observed broadband colors. 

There appears to be a dearth of stars with low chromospheric activity and high spot covering fractions. These observations are consistent with the solar perspective that plages are associated with large starspots, and plages can appear without starspots. 

Overall however, we caution users of the TiO absorption technique that moderate signal-to-noise spectra at $R\approx 31,500$ provide only weak detections of TiO absorption due to starspots. Using both higher resolution and higher signal-to-noise spectra in addition to light curve modelling \citep[as in][for example]{Gully2017} is preferable for robust spot covering fractions. 

\acknowledgements 

We express gratitude to James Davenport for helpful conversations during the preparation of this manuscript. We acknowledge the fantastic support staff at APO who made these observations possible: Russet McMillan, Candace Gray, Jack Dembicky and Ted Rudyk. 

J.L.C. acknowledges support provided by the NSF through grant AST-1602662.

Based on observations obtained with the Apache Point Observatory 3.5-meter and ARCSAT telescopes, which are owned and operated by the Astrophysical Research Consortium. This work presents results from the European Space Agency (ESA) space mission Gaia. Gaia data are being processed by the Gaia Data Processing and Analysis Consortium (DPAC). Funding for the DPAC is provided by national institutions, in particular the institutions participating in the Gaia MultiLateral Agreement (MLA). The Gaia mission website is \url{https://www.cosmos.esa.int/gaia}. The Gaia archive website is \url{https://archives.esac.esa.int/gaia}. This research has made use of the VizieR catalogue access tool, CDS, Strasbourg, France. The original description of the VizieR service was published in A\&AS 143, 23. This research has made use of NASA's Astrophysics Data System. This research has made use of the SVO Filter Profile Service (\url{http://svo2.cab.inta-csic.es/theory/fps/}) supported from the Spanish MINECO through grant AyA2014-55216 \citep{rodrigo2012}.

\software{\texttt{astroplan} \citep{astroplan}, \texttt{emcee}, \citep{Foreman-Mackey2013},\texttt{astropy} \citep{Astropy2018}, \texttt{ipython} \citep{ipython}, \texttt{numpy} \citep{VanDerWalt2011}, \texttt{scipy} \citep{scipy},  \texttt{matplotlib} \citep{matplotlib}, \texttt{SME} \citep{Valenti2005}, \moog \citep{moog}, \texttt{aesop} \citep{aesop}, \texttt{arcesetc} \citep{Morris2019}}

\facilities{APO/ARC 3.5 m Telescope, ESA/Gaia}

\vspace{2cm}


\appendix

\section{Stellar properties} \label{app:stellarprops}

In the following subsections, we will enumerate some references to stars in the sample. 

\subsection{70 Oph B (Gl 702 B)}

The smaller component in a nearby binary system (5 pc) 
discovered by \citet{Herschel1782}, 70 Ophiuchi B (Gl 702 B) is a $0.73 M_\odot$ star with $T_\mathrm{eff} = 4390 \pm 200$ K \citep{Eggenberger2008}. 

\subsection{51 Peg}
51 Peg hosts a hot Jupiter \citep{Mayor1995}.
\citet{Henry1997} measured a small upper limit on the 
amplitude of the rotational modulation of $\Delta(b~+~y)/2 < 0.0002 \pm 0.0002$ mag. 
\citet{Mittag2016} estimated the age of 51 Peg at 
$6.1 \pm 0.6$ Gyr, which may explain why it is significantly less active than the Sun \citep[e.g.,][]{Mamajek2008}.
 
\subsection{$\sigma$ Draconis}
This K0 star has a radial velocity periodicity that correlates with activity, indicating a stellar activity cycle with period $P_\mathrm{cyc} = 6$ years \citep{Butler2017}. 
It is bright and near enough ($G = 4.38$, 5.8 pc) 
that if it has large spots, 
they could be detectable as astrometric jitter in \gaia observations \citep{Morris2018e}.

\subsection{HD 39587 ($\chi^1$ Ori)}

HD 39587 is a spectroscopic binary with $M_A = 1.01 \pm 0.13 M_\odot$, and $M_B = 0.15 \pm 0.02 M_\odot$ \citep{Konig2002}, and the primary star has a rotation period $P_\mathrm{rot} = 5.36$ d \citep{Wright2011}. 
One can draw interesting comparisons to the young Sun, 
since the semimajor axis is 4.33 AU and the stars are 
unlikely to be tidally interacting.

\subsection{HD 68017}
HD 68017 is a G4V $M=0.85^{+0.04}_{-0.03} M_\odot$ star with an M5 companion 
at $\approx0\farcs6$, which at 21.8 pc translates to a projected 
separation of 13 AU \citep{Crepp2012}. It is likely that our spectrum contains both the primary and secondary stars, but the secondary star's flux at 7000 \angstrom should be 4\% of the primary star's flux, and so we treat the system as a single star for the purposes of starspot modeling, which has little effect on our interpretation of the overall results  (as we will find that the apparent spot covering fraction is consistent with zero).

\subsection{HD 110833}

HD 110833 is a K3+K3 binary with orbital period $P_\mathrm{orb} = 271$ d, and a tertiary mid-M dwarf 1\farcs6 away \citep{Rodriguez2015}.

\subsection{HD 127506}

HD 127506 hosts a non-transiting substellar companion with orbital period $P_\mathrm{orb}=7.1$ years discovered via radial velocities \citep{Reffert2011}. 

\subsection{HD 87884}

The K2V star HD 87884 ($\alpha$ Leo B) is remarkable for being the primary with a faint M4V companion, the pair of which may orbit Regulus \citep{Eggen1982,McAlister2005,Jankov2017}. The faint secondary should contribute $\lesssim 10\%$ of the flux of the primary at 7000 \angstrom, so we note with some caution that the small, but non-zero coverage in this system might arise from significant TiO absorption in the spectrum of the secondary component. 

\subsection{HD 79555}

This young (35 Myr) K4V star is in a long period astrometric binary separated by 0.7$''$ in the Castor moving group \citep{Mason2001, Maldonado2010}. 

\subsection{HD 73667}

This star was identified by \citet{Wright2004b} as a potential Maunder minimum star 
\citep{Eddy1976} for its low 
chromospheric activity ($\logrprime = -4.97$ dex).

\subsection{GJ 9781 A}

This K7 star has an astrometric binary companion with separation 147\farcs8 and $\Delta V = 9.13$, which is easily separated in our ARCES spectroscopy \citep{Lepine2007}. 

\section{Color combinations}

From \citet{Neff1995}, the composite flux $F _ { \text { total } }$ from a linear combination a quiescent spectrum $F_Q$ and a spotted spectrum $F_S$, with spot covering fraction $f_S$ is
\begin{equation}
F _ { \text { total } } = \frac { f _ { S } R _ { \lambda } F _ { S } + \left( 1 - f _ { S } \right) F _ { Q } } { f _ { S } R _ { \lambda } + \left( 1 - f _ { S } \right) }, 
\end{equation}
where $R_\lambda$ is the ratio of continuum fluxes of the spotted and unspotted spectra at wavelength $\lambda$. Thus the linear weighting functions for the quiescent and spotted components respectively are:
\begin{eqnarray}
W _ { Q } &=& \left( 1 - f _ { S } \right) / \left[ \left( f _ { S } R _ { \lambda } \right) + \left( 1 - f _ { S } \right) \right], \\ 
W _ { S } &=& \left( f _ { S } R _ { \lambda } \right) / \left[ \left( f _ { S } R _ { \lambda } \right) + \left( 1 - f _ { S } \right) \right].
\end{eqnarray}
Therefore the composite apparent magnitude of a star, composed of a combination of atmospheres with two different colors $c_Q$ and $c_S$, is given by applying the linear weighting in flux space and transforming back to magnitude space: 
\begin{equation}
m = 2.5 \log_{10}\left( W_Q 10^{c_Q/2.5} + W_S 10^{c_S/2.5} \right). \label{eqn:color}
\end{equation}

\section{\moog results} \label{sec:moogresults}

For details on how these results were derived, see Section~\ref{sec:moog}. 

\startlongtable
\begin{deluxetable}{cccccc}
\tablehead{\colhead{Target} & \colhead{$T_\mathrm{eff, MOOG}$} & \colhead{$\log g$} & \colhead{Micro.} & \colhead{[\ion{Fe}{1}/H]} & \colhead{[\ion{Fe}{2}/H]}}
\startdata
51 Peg & $5800 \pm 115$ & $4.42 \pm 0.05$ & $1.05 \pm 0.17$ & $0.14 \pm 0.02$ & $-0.07 \pm 0.10$ \\
GJ 702 B & $4385 \pm 65$ & $4.64 \pm 0.05$ & $1.01 \pm 0.13$ & $0.29 \pm 0.02$ & $0.68 \pm 0.02$ \\
GJ 9781 A & $4500 \pm 55$ & $4.63 \pm 0.05$ & $0.97 \pm 0.10$ & $0.07 \pm 0.02$ & $0.57 \pm 0.17$ \\
HAT-P-11 & $4790 \pm 175$ & $4.61 \pm 0.05$ & $1.03 \pm 0.29$ & $-0.06 \pm 0.02$ & $0.21 \pm 0.10$ \\
HD 10697 & $5480 \pm 35$ & $4.51 \pm 0.10$ & $0.93 \pm 0.10$ & $-0.05 \pm 0.01$ & $0.0 \pm 0.12$ \\
HD 110833 & $5130 \pm 85$ & $4.54 \pm 0.05$ & $1.17 \pm 0.10$ & $0.08 \pm 0.01$ & $-0.12 \pm 0.11$ \\
HD 113827 & $4510 \pm 65$ & $4.63 \pm 0.05$ & $1.15 \pm 0.10$ & $-0.14 \pm 0.02$ & $-0.07 \pm 0.15$ \\
HD 122120 & $4960 \pm 65$ & $4.6 \pm 0.05$ & $1.89 \pm 0.36$ & $-0.20 \pm 0.02$ &  \\
HD 127506 & $4955 \pm 70$ & $4.6 \pm 0.05$ & $1.38 \pm 0.22$ & $-0.10 \pm 0.01$ & $-0.27 \pm 0.10$ \\
HD 129333 & $5490 \pm 40$ & $4.51 \pm 0.05$ & $2.07 \pm 0.10$ & $-0.15 \pm 0.01$ & $0.22 \pm 0.03$ \\
HD 134319 & $5640 \pm 70$ & $4.47 \pm 0.05$ & $1.47 \pm 0.11$ & $-0.08 \pm 0.02$ & $0.01 \pm 0.10$ \\
HD 148467 & $4500 \pm 75$ & $4.63 \pm 0.05$ & $1.08 \pm 0.10$ & $-0.10 \pm 0.02$ &  \\
HD 151288 & $4395 \pm 90$ & $4.65 \pm 0.05$ & $1.8 \pm 0.52$ & $-0.32 \pm 0.03$ &  \\
HD 175742 & $4930 \pm 85$ & $4.6 \pm 0.05$ & $1.57 \pm 0.39$ & $0.04 \pm 0.02$ & $0.20 \pm 0.09$ \\
HD 182488 & $5320 \pm 65$ & $4.54 \pm 0.05$ & $0.77 \pm 0.10$ & $0.16 \pm 0.01$ & $0.09 \pm 0.08$ \\
HD 200560 & $4800 \pm 105$ & $4.61 \pm 0.05$ & $1.04 \pm 0.12$ & $0.08 \pm 0.01$ & $0.15 \pm 0.10$ \\
HD 209290 & $4020 \pm 50$ & $4.73 \pm 0.05$ & $0.71 \pm 0.40$ & $-0.46 \pm 0.02$ &  \\
HD 210277 & $5490 \pm 30$ & $4.51 \pm 0.05$ & $0.90 \pm 0.29$ & $0.24 \pm 0.01$ & $0.10 \pm 0.06$ \\
HD 220182 & $5090 \pm 65$ & $4.58 \pm 0.05$ & $0.88 \pm 0.10$ & $-0.03 \pm 0.01$ & $0.18 \pm 0.07$ \\
HD 221639 & $4900 \pm 40$ & $4.6 \pm 0.05$ & $0.94 \pm 0.14$ & $0.33 \pm 0.01$ & $0.54 \pm 0.10$ \\
HD 222107 & $4590 \pm 30$ & $4.58 \pm 0.05$ & $0.97 \pm 0.40$ & $-0.19 \pm 0.01$ &  \\
HD 266611 & $4230 \pm 40$ & $4.67 \pm 0.05$ & $0.79 \pm 0.38$ & $-0.26 \pm 0.02$ &  \\
HD 34411 & $5920 \pm 50$ & $4.36 \pm 0.05$ & $1.34 \pm 0.15$ & $0.02 \pm 0.01$ &  \\
HD 38230 & $5130 \pm 70$ & $4.58 \pm 0.05$ & $0.93 \pm 0.35$ & $-0.24 \pm 0.01$ & $-0.15 \pm 0.23$ \\
HD 39587 & $5800 \pm 110$ & $4.42 \pm 0.05$ & $1.16 \pm 0.10$ & $-0.15 \pm 0.01$ & $-0.03 \pm 0.07$ \\
HD 41593 & $5040 \pm 75$ & $4.58 \pm 0.05$ & $0.76 \pm 0.16$ & $-0.06 \pm 0.01$ & $0.07 \pm 0.11$ \\
HD 42250 & $5400 \pm 45$ & $4.53 \pm 0.05$ & $0.99 \pm 0.12$ & $-0.01 \pm 0.01$ & $-0.11 \pm 0.09$ \\
HD 47752 & $4800 \pm 30$ & $4.61 \pm 0.05$ & $0.53 \pm 0.22$ & $-0.07 \pm 0.01$ & $-0.14 \pm 0.10$ \\
HD 50692 & $5930 \pm 50$ & $4.39 \pm 0.05$ & $1.21 \pm 0.10$ & $-0.26 \pm 0.01$ & $-0.44 \pm 0.07$ \\
HD 5857 & $4820 \pm 30$ & $3.44 \pm 0.10$ & $1.81 \pm 0.10$ & $-0.12 \pm 0.01$ & $-0.14 \pm 0.10$ \\
HD 62613 & $5430 \pm 50$ & $4.53 \pm 0.05$ & $0.93 \pm 0.10$ & $-0.14 \pm 0.01$ & $-0.35 \pm 0.11$ \\
HD 6497 & $4510 \pm 80$ & $2.37 \pm 0.10$ & $1.41 \pm 0.10$ & $-0.11 \pm 0.05$ & $-0.18 \pm 0.10$ \\
HD 68017 & $5690 \pm 45$ & $4.49 \pm 0.05$ & $0.99 \pm 0.10$ & $-0.39 \pm 0.01$ & $-0.55 \pm 0.06$ \\
HD 73667 & $5090 \pm 30$ & $4.62 \pm 0.05$ & $0.84 \pm 0.12$ & $-0.63 \pm 0.01$ & $-0.73 \pm 0.14$ \\
HD 86728 & $5790 \pm 50$ & $4.43 \pm 0.05$ & $0.99 \pm 0.10$ & $0.23 \pm 0.01$ & $0.10 \pm 0.10$ \\
HD 87884 & $4920 \pm 40$ & $4.6 \pm 0.05$ & $0.77 \pm 0.14$ & $-0.16 \pm 0.01$ & $-0.08 \pm 0.14$ \\
HD 88230 & $4310 \pm 50$ & $4.65 \pm 0.10$ & $0.24 \pm 0.10$ & $-0.13 \pm 0.01$ &  \\
HD 89744 & $6070 \pm 110$ & $4.18 \pm 0.05$ & $1.42 \pm 0.10$ & $0.04 \pm 0.01$ & $0.34 \pm 0.01$ \\
HD 98230 & $5580 \pm 70$ & $4.52 \pm 0.05$ & $0.8 \pm 0.12$ & $-0.41 \pm 0.02$ & $-0.13 \pm 0.05$ \\
$\sigma$ Draconis & $5130 \pm 40$ & $4.58 \pm 0.05$ & $0.73 \pm 0.10$ & $-0.24 \pm 0.01$ & $-0.08 \pm 0.08$ \\
\enddata
\end{deluxetable}

\end{document}